\documentclass[9pt, conference]{IEEEtran}
\IEEEoverridecommandlockouts

\usepackage{cite}
\usepackage{amsmath,amssymb,amsfonts}
\usepackage{algorithmic}
\usepackage[ruled,linesnumbered,vlined]{algorithm2e}
\usepackage{graphicx}
\usepackage{textcomp}
\usepackage{xcolor}
\usepackage{url}
\usepackage{xspace}
\usepackage{booktabs, multirow} %
\usepackage{listings}
\usepackage{subfig}
\usepackage{float}
\usepackage{textgreek}
\usepackage{multirow, makecell}
\usepackage{balance}
\usepackage[hidelinks,colorlinks=true,citecolor=teal,linkcolor=purple,urlcolor=black]{hyperref}
\usepackage[noabbrev,capitalise]{cleveref}
\usepackage{enumitem}
\usepackage{wrapfig}
\usepackage{stfloats}
\usepackage{xfrac}
\usepackage{comment}

\def\BibTeX{{\rm B\kern-.05em{\sc i\kern-.025em b}\kern-.08em
    T\kern-.1667em\lower.7ex\hbox{E}\kern-.125emX}}
\newcommand{\URL}[1]{\begin{small}\url{#1}\end{small}}

\newcommand{\Comment}[1]{}
\newcommand{\Space}[1]{}

\newcommand{\CodeIn}[1]{\texttt{#1}}
\newcommand{\ccc}[1]{\hfill \begin{footnotesize}\# #1\end{footnotesize}}

\newcommand{\pne}{NE\xspace}
\Crefname{figure}{Fig.}{Figs.}
\crefname{figure}{Fig.}{Figs.}
\crefname{algorithm}{Alg.}{Algs.}
\Crefname{algorithm}{Alg.}{Algs.}
\crefname{section}{\S}{\S}

\begin{document}

\title{Algorithmic Tradeoff Exploration for Component Placement and Wire Routing in Nanomodular Electronics}

\author{
\IEEEauthorblockN{Peidi Song$^*$\qquad Alexandros Daglis$^*$\qquad Michael Filler$^\dag$\qquad Ahmed Saeed$^*$}
\IEEEauthorblockA{\textit{$^*$School of Computer Science}\qquad \textit{$^\dag$School of Chemical and Biomolecular Engineering}}
\IEEEauthorblockA{\textit{Georgia Institute of Technology}}
}

\pagestyle{plain}

\maketitle

\begin{abstract}

Advances in fabrication technology have enabled modularizing electronic components at the micro- or nano-scale and composing these modules on demand into larger circuits.
Micromodular and nanomodular electronics (ME and \pne) open a new design space in electronics, promising a degree of flexibility, extensibility, and accessibility far superior to traditional monolithic methods.
ME/\pne leverage a multi-stage process of initial imprecise component deposition, followed by precise wire printing to compose them into a circuit.
Due to imperfections in deposition, each circuit instance has a unique layout with its own component placement and wire routing solutions, putting the design automation process on the critical path.
Moreover, high-performance nanomodular components enable the synthesis of larger heterogeneous circuits than traditional printed electronics, requiring more scalable algorithms.
ME/\pne thus introduce a tradeoff between the time-to-solution for placement/routing algorithms and the resulting total wire length, with the latter dictating circuit printing time.
We explore this tradeoff by adapting standard partitioning, floorplanning, placement, and routing algorithms to the unique characteristics of ME/\pne.
Our evaluations demonstrate significant optimization headroom in different dimensions.
For example, our tunable algorithms can deliver a $\mathbf{108\times}$ improvement in end-to-end manufacturing time at the cost of $\mathbf{21\%}$ increase in total wire length.
Conversely, circuit quality/performance can be prioritized at the cost of increased manufacturing time, highlighting the value of the ability to dynamically navigate the tradeoff space according to the primary optimization metric.

\end{abstract}

\section{Introduction}

Today's processes for manufacturing microelectronics involve a tradeoff between circuit performance and customization. 
While integrating millions or billions of components into a single, general-purpose, high-performance circuit is a modern marvel, it is still challenging to integrate materials with different physical properties into one circuit or %
customize hardware except in the most high-value scenarios. This tradeoff is a result of the 70-year-old planar process, which, while world-changing, requires per-step perfection in terms of manufacturing, which drives centralized production, high capital costs, and long end-to-end process times.

Micromodular and nanomodular electronics (ME and \pne) are an alternative approach that distributes component fabrication and circuit wiring in space and time, aiming to enable {extreme customization}, {deep heterogeneous integration}, and {on-demand manufacturing}. The process begins by fabricating high-performance components. Some of these components are nanomodular, incorporating individual transistors or memristors, while others are micromodular, containing hundreds or thousands of individual elements for computing, memory, power harvesting, or other advanced functions. Once fabricated, these component modules are deposited and interconnected through high-resolution printing~\cite{filler2023nanomodular}.
This capability unlocks a new design space for electronics, creating opportunities for use in highly size- and power-constrained applications, such as wearable, ingestible, or implantable healthcare devices. Furthermore, ME/\pne dramatically lower the barrier to entry, democratizing custom electronics. For brevity, we henceforth use the single term ``NE'' to refer to both ME and its technological evolution, NE.

A key challenge in \pne is that manipulating and placing micro/ nanoscale modules differs significantly from handling larger-scale components, as pick-and-place methods do not easily scale to such component dimensions. Self-assembly theoretically offers the necessary scalability, but suffers from a time/precision tradeoff. Rather than pursuing complex and time-consuming techniques to approach perfect component deposition, which becomes ever more challenging as components shrink, \pne demand a different approach that shifts a complex manufacturing challenge into a computational problem, potentially enhancing the robustness and simplifying the manufacturing process.
\pne circuit manufacturing comprises the following steps: 

\noindent \textbf{1) Component Fabrication.} Nanomodular or micromodular components are fabricated using traditional fab-based processes or newer bottom-up methods. Component ``inks" are created by suspending these components in a solution.

\noindent \textbf{2) Component Deposition.} One or more component inks are mixed and deposited onto a substrate. This component deposition process involves a speed-precision tradeoff, resulting in imprecise component positions and orientations. In some situations, deposition may result in entirely random component locations. 
 
\noindent \textbf{3) Deposition Analysis, Logical-to-Physical Component Assignment (Placement), and Wire Planning.} A precise vision system analyzes component positions and orientations~\cite{potocnik2022automated}. The target circuit's logical components are then assigned physical components on the substrate, and wiring paths are planned to interconnect the physical components to compose the intended logical circuit.

\noindent \textbf{4) Wire Printing.} Conductive wires (and insulators) are precisely printed among the nanomodular components via e-jet~\cite{park2007high, galliker2012direct, barton2010desktop}, following the wiring plan generated in the previous step. 

In this paper, we focus on the third step.
We consider the problem of component placement and wire routing for a logical circuit over a given layout of physical components. \pne introduce unique challenges to these problems. 
First, \pne put \textit{design automation on the critical path} of the manufacturing process of every physical circuit. In particular, placement and routing must be performed separately for each physical circuit due to imprecise component deposition. Traditional manufacturing amortizes design automation costs over millions of units, but this approach is infeasible in countless low-volume situations. Hence, it is necessary to balance the overhead of generating a high-quality solution with the overall wire printing time, which represents a considerable fraction of the end-to-end manufacturing process.

Second, \pne introduce \textit{unique constraints on legal component placements}. In particular, \pne share most of the constraints on placement (e.g., avoiding overlapping components and creating routable placement solutions), but also requires that individual logical components are mapped to pre-positioned physical components. In other words, while placement decisions in most existing scenarios consider space as continuous and uniform, it is discrete and non-uniform in \pne. %

Third, routing in \pne must \textit{scale to large circuits without compromising time-to-solution}, as a high overhead to identify an optimal solution does not get amortized over large circuit volume. While our formulation of the routing problem resembles routing in printed electronics~\cite{wiklund2021review}, there is a significant difference in scale. \pne compose hundreds of thousands of high-performance components, rather than %
hundreds of low-performance components.

Our goal is to explore the tradeoff space between %
\textit{time-to-solution} and \textit{solution quality}, subject to the unique constraints of \pne, by adapting standard partitioning, floorplanning, placement, and routing algorithms to the unique constraints of \pne. In particular, we use simulated annealing to solve the placement problem. Further, we employ BFS to route components within a circuit partition, and weighted A* to route between partitions. 
We focus on the standard objective of reducing the total wire length used in a circuit~\cite{lu2015eplace, gu2020dreamplace, lin2022gamer, lin2022superfast}, which remains relevant in \pne, as wire-printing time is a major determinant of the total manufacturing time.

In summary, we make the following contributions:
\begin{itemize}[noitemsep,topsep=0pt,leftmargin=*]
    \item We are the first to formulate the component placement and wire routing problem for \pne, identifying unique constraints and the tradeoff space between time-to-solution and solution quality.
    \item We present \pne design automation workflow and adapt placement and routing algorithms that capture design space tradeoffs and balance all requirements.
    \item We conduct an evaluation demonstrating how our suite of algorithms enables selective movement in the tradeoff space, based on the target optimization metric. For example, we show an opportunity for $108\times$ improvement in end-to-end manufacturing time at the cost of $21\%$ increase in the target circuit's total wire length.
    
\end{itemize}

\section{Nanomodular Electronics Background}

\pne factor micro/nanomodular component fabrication and circuit wiring in space and time, in contrast to the entirely integrated nature of today's fabs. %
This process combines the component performance benefits of custom microelectronics with the manufacturing speed and heterogeneous component composition capability of printed circuit boards. 
Components are high-performance because they benefit from traditional cleanroom methods. For example, Elly et al. show that high-performance micromodular silicon transistors can be transferred to non-native substrates and interconnected to form functional nMOS inverters, achieving near-wafer electron mobilities and high gain~\cite{10684980}. Notably, the performance metrics demonstrated by Elly et al. are at least an order of magnitude superior to state-of-the-art fully printed transistors~\cite{molina2019inkjet}. \pne also allow for the integration of a diversity of materials and device types, enabling memory, logic, and interconnect elements to be incorporated side by side. Kurup et al. demonstrate modular HfOx-based resistive random access memory (RRAM) with low-voltage multi-level switching, illustrating how micromodular memory can expand the functional palette of \pne circuits~\cite{kurup:modular}.

While component deposition is imprecise (as discussed next), precise wiring can be accomplished with high-resolution methods such as electrohydrodynamic jet (e-jet) printing, achieving sub-micron line resolutions and print speeds over 1 cm/s~\cite{park2007high, galliker2012direct, barton2010desktop}. Methods like e-jet also exhibit footprints smaller than a desktop, ensuring portability and enhancing manufacturing security and cost efficiency, representing a significant departure from the multi-billion-dollar centralized fabs. The feasibility of this approach is demonstrated by Yue et al., who develop an adaptive, process-aware desktop wiring framework using e-jet printing. Their experiments show reliable routing of micromodular systems with submicron interconnect resolution and high device yield, establishing a clear experimental foundation for \pne wiring~\cite{yue_automated_2025}.

There are broader manufacturing and economic benefits as well. In traditional manufacturing, non-recurring engineering costs are spread over millions of units, which excludes use cases that lack the production scale to absorb these costs. In \pne, however, component costs can be distributed across multiple microsystems, rather than depending on a single microsystem to achieve sufficient scale. Using prefabricated modular components and additive wiring methods significantly reduces microsystem manufacturing time, enabling circuits to be better tailored to their workload.

However, \pne's multi-step manufacturing process introduces new challenges when placing components on the substrate.
Scaling component dimensions down to the microscale and below poses significant challenges for traditional pick-and-place deposition methods due to two main factors: (1) deposition rate and (2) manipulation difficulty. 
There is often a tradeoff between deposition precision and process speed \cite{zhou2020device, freer2010high}.
Even in the most meticulously controlled and slow processes, achieving near-perfect deposition is rare.

Consider a 1 mm$^2$ microsystem that, although monolithic and manageable with today’s pick-and-place deposition technology, may in the future be constructed from 100 µm$^2$ heterogeneous micromodular components. Assembling this same 1 mm$^2$ microsystem would require picking and placing 10k individual components—demanding a 10k-fold increase in pick-and-place speed. While advancements in this area are expected, achieving a four-orders-of-magnitude improvement will be a considerable challenge. Furthermore, the required rate of increase grows as components continue to scale down. Separate from manufacturing rate, the physics of microscale and smaller objects makes the deterministic manipulation inherently challenging. Small objects are more strongly affected by thermal fluctuations, are more likely to adhere to surfaces due to the relative strength of van der Waals forces at these scales, and have limited controllability over the direction of interactions. %
These limitations motivate our interest in microsystems built with imperfect and even random deposition of micro/nanomodular components. Importantly, although component deposition may be random, modern automation and additive manufacturing tools enable precise wire printing, allowing for corrections to account for deposition imperfections.

\section{Scope and Related Work}

The reality of imprecise component deposition in \pne implies the need to generate a distinct physical design per manufactured circuit, placing the design automation process on the critical path, instead of a one-time initial overhead.
The necessity of generating solutions \textit{per physical chip} rather than per logical design urges the design of algorithms that \textit{co-optimize solution quality with time-to-solution}.  

We approach the circuit design problem in two consecutive steps: \textit{component placement} maps logical components to physical components on the substrate; \textit{wire routing} plans wires and insulation materials on the substrate given the component placement result. In this section, we highlight the unique challenges posed by \pne compared to other electronics manufacturing processes.

\smallskip
\noindent \textbf{Design automation optimization goal: balancing solution quality with speed.}
We focus on the standard objective of reducing the total wire length, which directly affects the quality of the produced solution and circuit printing time, as wires must be sequentially printed.
At the same time, given imprecise component deposition, a placement and routing solution must be produced per circuit rather than per logical design, placing the time-to-solution on the critical path of each circuit's manufacturing.
Hence, our design automation algorithms must balance the quality of the produced placement and routing solution in terms of resulting wire length with the time required to produce that solution.

\smallskip
\noindent \textbf{Restricted legal placements.} \pne share most of the constraints on legal placements as existing electronics manufacturing techniques (e.g., avoiding overlapping components and creating routable placement solutions). 
To appreciate the added complexity, conventional electronics applying analytical placement in a continuous space perform minor legalization steps afterwards to avoid overlaps and align circuit cells with channels, where individual channels are still continuous~\cite{spindler2008abacus, pentapati2023legalization}.
Placement in FPGAs forces space to be discrete as circuit cells need to be assigned to individual blocks on board, complicating the legalization step~\cite{gort2012analytical, bian2010towards}. However, FPGA blocks tend to be uniformly placed. 
In contrast, placement in \pne corresponds to assigning individual logical components to individual prepositioned physical components to produce a routable valid solution. 
Imperfectly deposited components create a \textit{discrete and non-uniform} space, further complicating any legalization step. 
While other standard approaches for placement and legalization require either continuous or uniform space, stochastic approaches like simulated annealing directly provide discrete placement decisions, with the ability to converge to a nearly optimal solution in a non-uniform exploration space.
Thus, in this paper, we focus on simulated annealing-based placement algorithms~\cite{seemuth2015automatic, mao2016modular}, adapting them to \pne requirements by unifying legalization with placement. However, simulated annealing is known to be slow. %
We explore techniques that make minor sacrifices in placement quality to make significant gains in terms of time-to-solution.

Recent works leverage reinforcement learning and machine learning to solve the placement problem~\cite{goldie2020placement, liu2022xplace, lu2023dream, cheng2023assessment}. %
We find our developed stochastic methods lightweight and effective enough for our purposes and leave applying such approaches to future work.

\smallskip
\noindent \textbf{Fast routing of intersecting wires.} \pne share routing challenges with printed electronics. Both processes connect components through wires that are laid on a single plane, allowing wires to intersect with insulation inserted at the intersection point~\cite{sanchez2008inkjet}.
State-of-the-art approaches for routing in printed electronics do not optimize for time-to-solution and can take several hours to route one single design with as small as hundreds of nets~\cite{rasheed2019predictive, rasheed2020crossover}.
In contrast, for \pne, we aim to produce solutions for every large circuit instance within minutes.

\section{Approach Overview}
\label{sec:approach overview}

\begin{figure*}[t]
    \centering
    \subfloat[NAND gate circuit.] {\label{fig:formulation/circuit_design}\includegraphics[width=0.5\columnwidth]{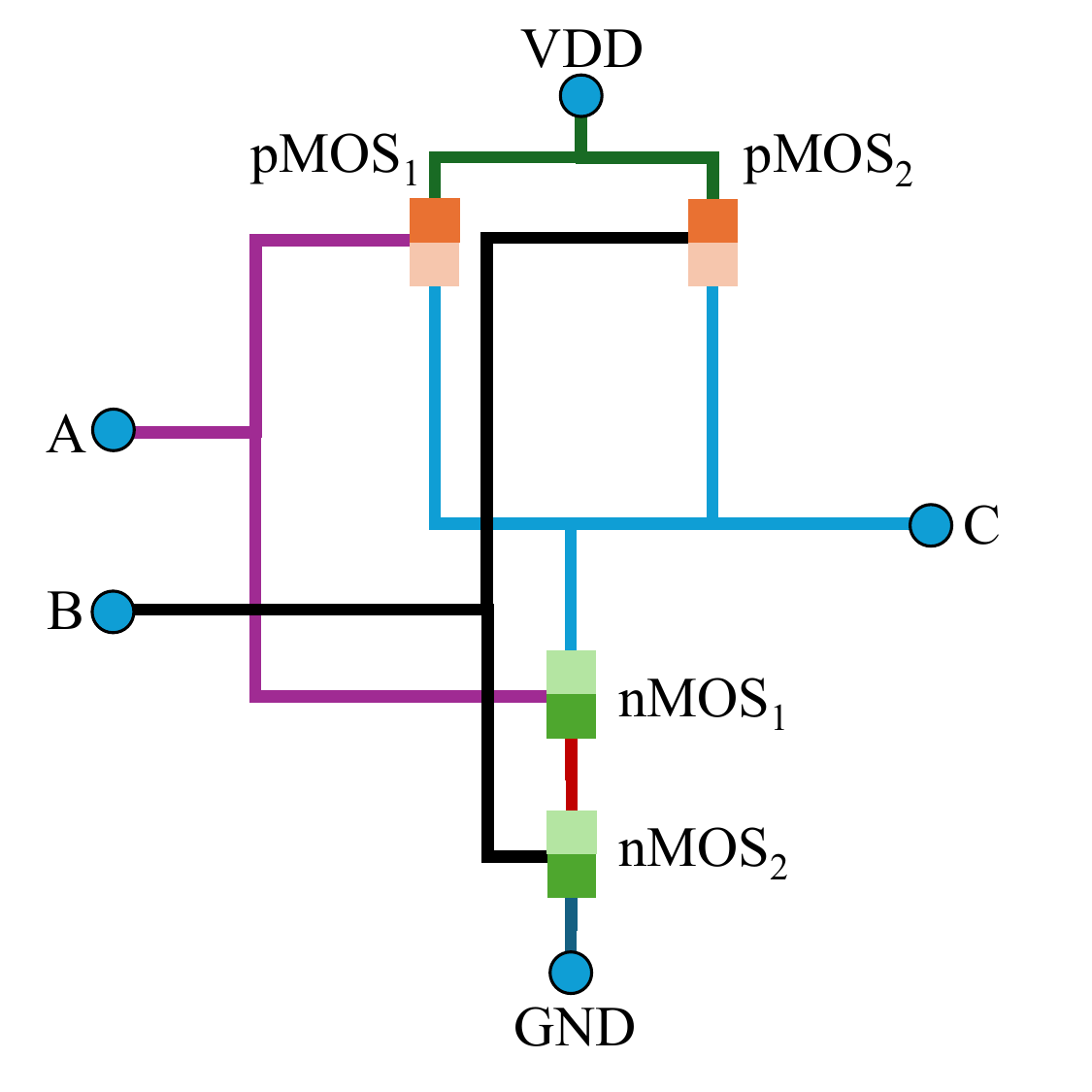}}
    \subfloat[Physical component layout.] {\label{fig:formulation/component_image}\includegraphics[width=0.5\columnwidth]{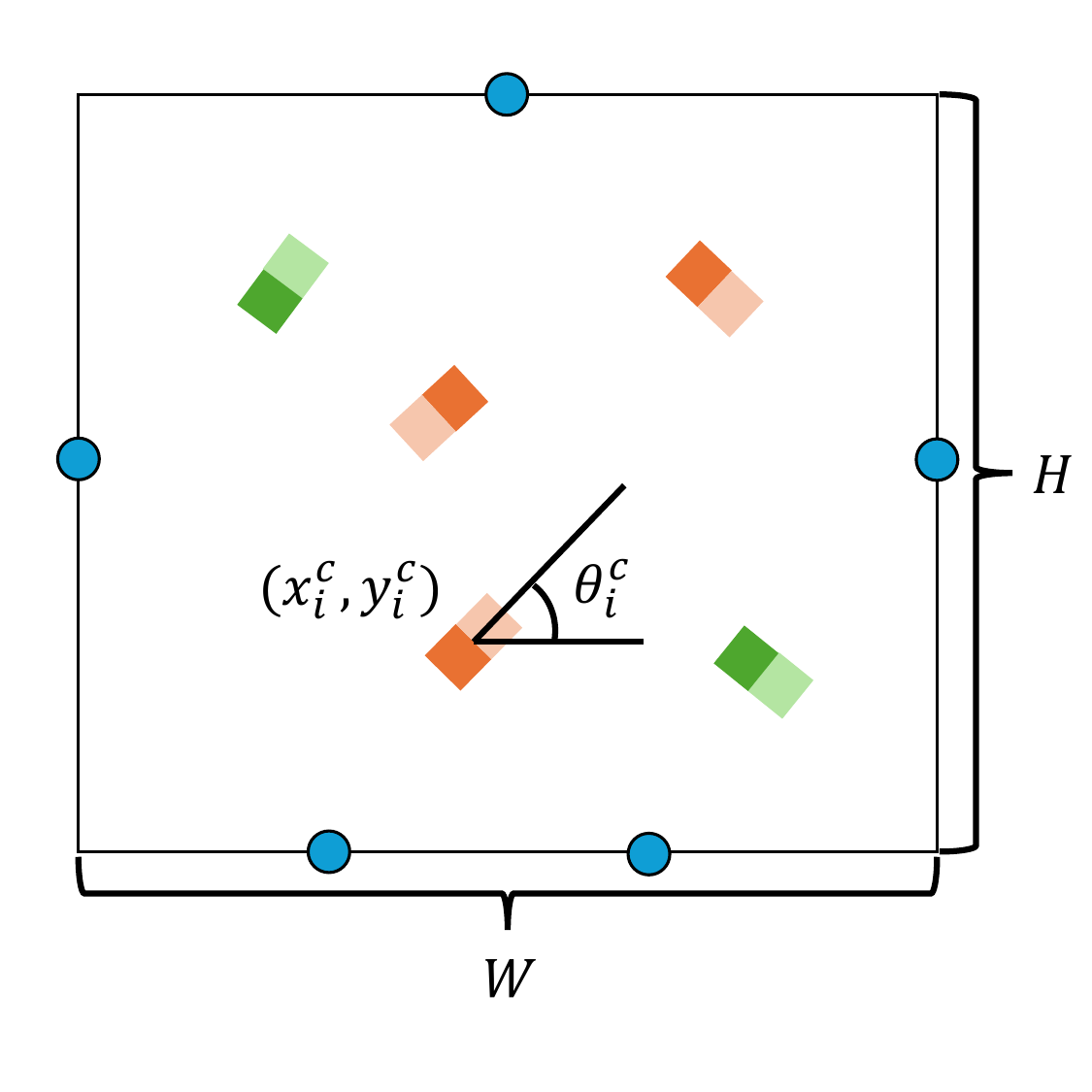}}
    \subfloat[Component placement solution.] {\label{fig:formulation/placement_result}\includegraphics[width=0.5\columnwidth]{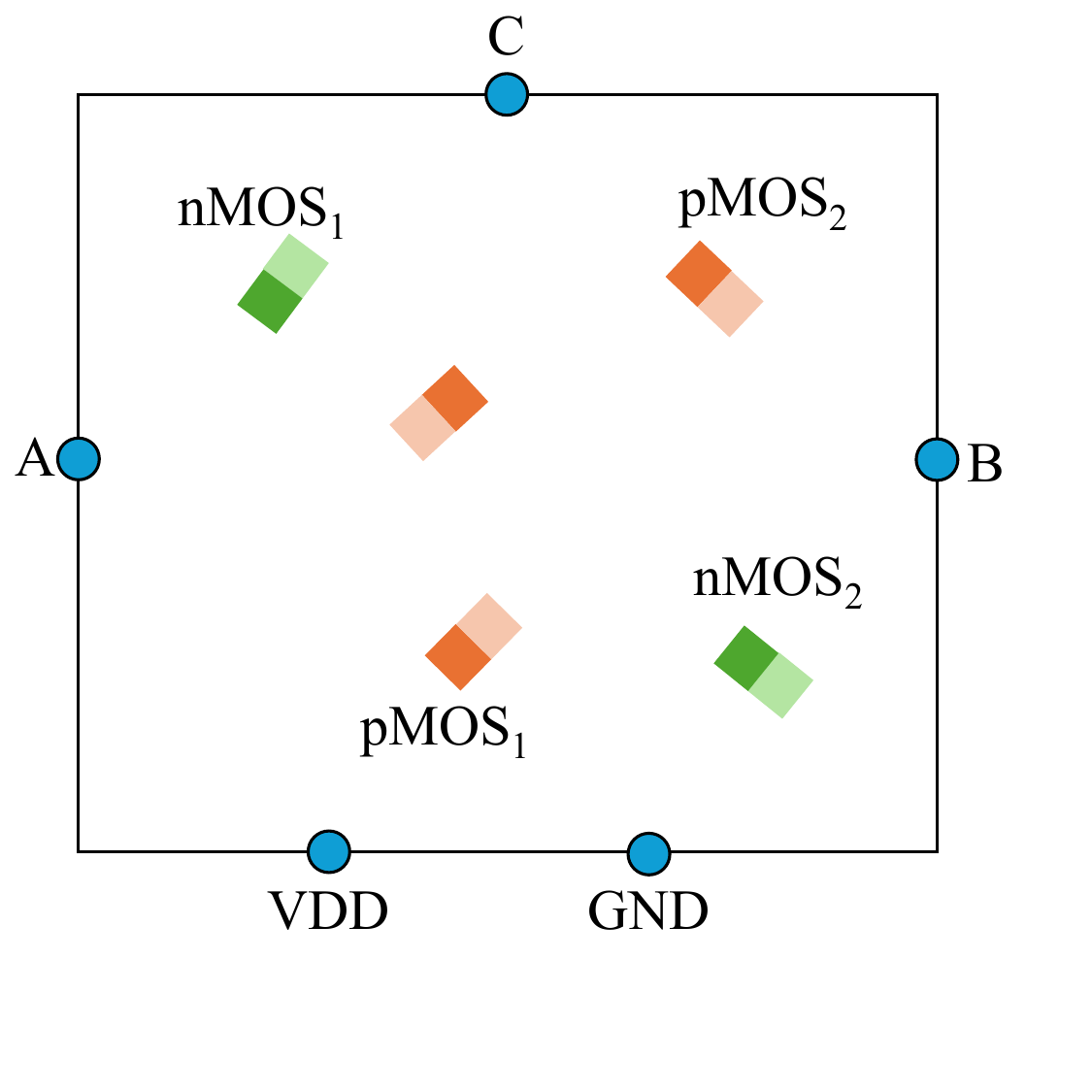}}
    \subfloat[Wire routing solution.] {\label{fig:formulation/routing_result}\includegraphics[width=0.5\columnwidth]{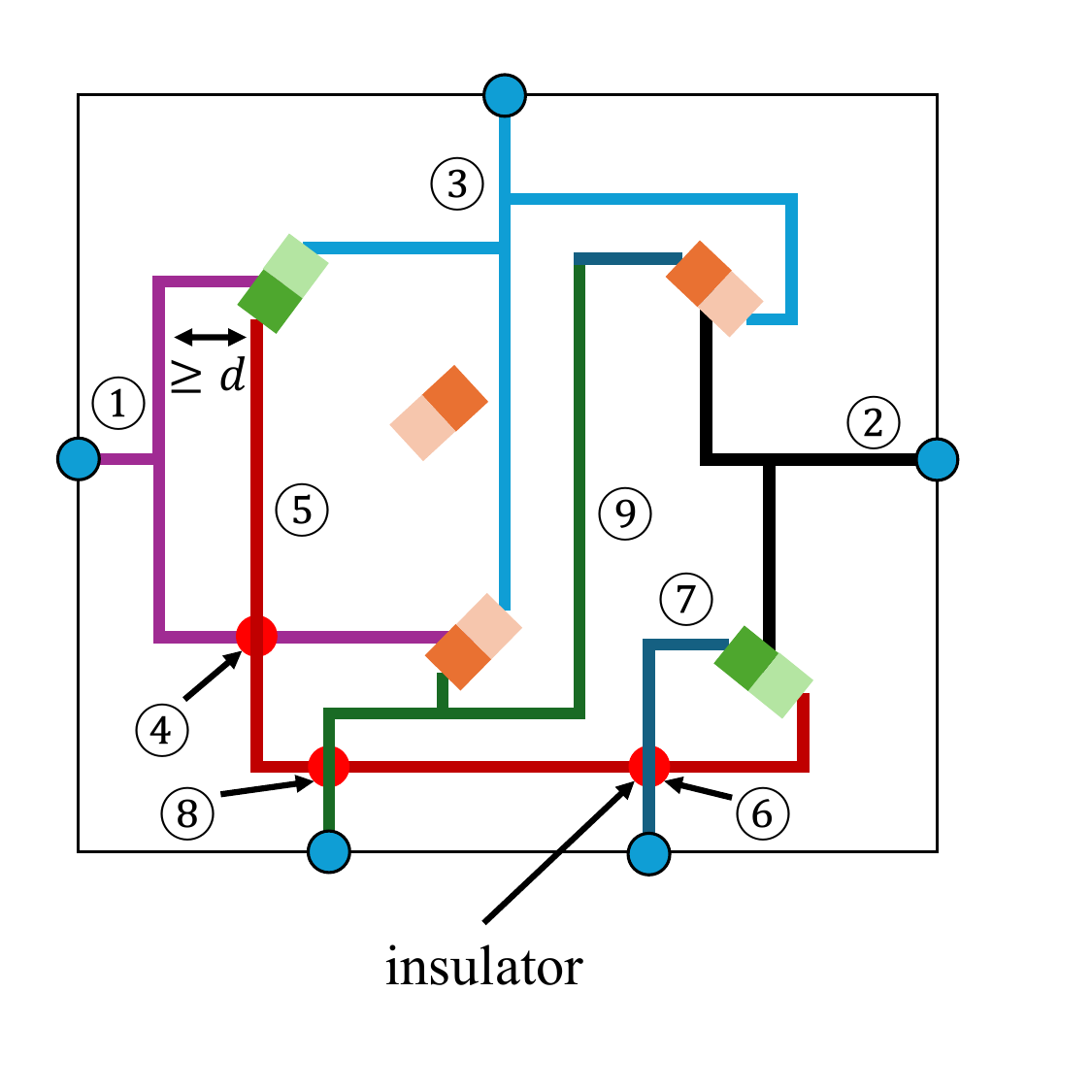}}
    \caption{Problem formulation. Inputs: (a) and (b). Outputs: (c) and (d). Numbers in (d) represent print order.}
    \label{fig:formulation}
\end{figure*}

\begingroup

\cref{fig:formulation} shows an example of the inputs and outputs of the placement and routing algorithms in \pne. Inputs include a transistor-level \textit{logical circuit design} (\cref{fig:formulation/circuit_design}), and a \textit{physical component layout} (\cref{fig:formulation/component_image}) %
derived from imaging the deposited components. 
We focus on transistor-level designs, but our algorithms can be extended to support additional components, such as memristors, or even higher-level components, such as gates or more complex modules.
We consider logical circuits with $N^L$ transistors with 3 pins per transistor, $S$ I/O pins and $E$ nets. 
Physical components are deposited on a 2D $W\times H$ rectangular substrate with $N^P$ physical components, and $S'$ I/O slots on the boundary. 
The distribution of components forms an approximate mesh---components are intended to be deposited onto the substrate forming a 2D array, while deposition imprecision alters the components' intended position and orientation.
The component size and density on state-of-the-art fabrication techniques~\cite{filler2023nanomodular} results in much larger average distance between components than the transistor length, making for a sparse space.
We assume $N^P\approx 2N^L$, introducing redundancy to mitigate the impact of potentially defective components~\cite{filler2023nanomodular}. 
The output of the \textit{placement} algorithm (\cref{fig:formulation/placement_result}) is a mapping of individual logical components and I/O pins to unique physical components and I/O slots. The output of the \textit{routing} algorithm (\cref{fig:formulation/routing_result}) is the layout of all wires and insulators to be printed and their printing order. Any two pins in the same net are connected by wires, and wires in different nets either do not interfere (distance $\geq d$) or are insulated at their intersection point.
We defer formal notations for inputs and outputs to \cref{sec:formalization}.

Our objective is to minimize the total print time, which is dominated by the total wire length ($\Psi$), and the overhead of computing placement and routing solutions for every printed circuit instance. The time overhead of printing insulators is eliminated by employing two printing nozzles: one for conductors and another trailing nozzle for insulators. In our evaluation, we report both $\Psi$ and the number of printed insulators ($\Omega$) to provide a complete picture of printing overheads.

\setlength{\intextsep}{0pt}%
\setlength{\columnsep}{8pt}%
\begin{wrapfigure}[18]{R}{0.5\columnwidth}
    \centering
    \includegraphics[width=0.5\columnwidth]{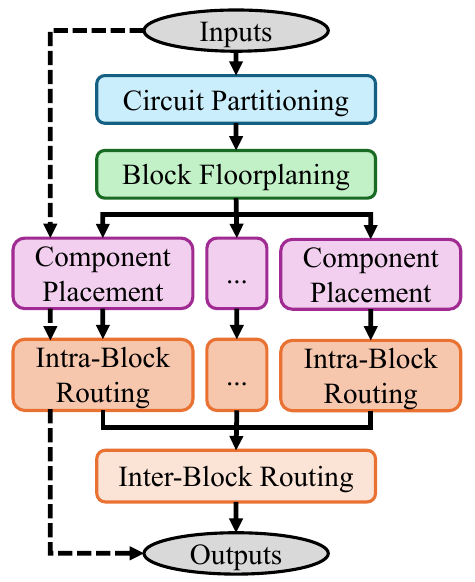}
    \caption{Our workflow for \pne design automation. Dashed arrows represent bypass for small circuits.}    
    \label{fig:new_workflow}
\end{wrapfigure}

\cref{fig:new_workflow} shows our \pne design automation workflow.
We follow a workflow comprising four distinct steps: circuit partitioning, block floorplanning, component placement, and routing, where routing is separated into two substeps of intra-block routing and inter-block routing.
Component placement and intra-block routing are performed in parallel across blocks.
When the circuit is too small to partition, we apply a bypassed workflow with only component placement and intra-block routing.
We introduce our approach for placement and routing in \cref{sec: SA comp assign} and \cref{section:routing}, respectively.

For circuit partitioning, we use the multi-way Fiduccia-Mattheyses (FM) algorithm to evenly partition the circuit into $B$ blocks~\cite{sanchis1989multiple}.
FM minimizes the cut size (i.e., number of nets spanning multiple blocks).
By assigning a higher cost to nets straddling more than two blocks, FM reduces the number of such nets, thus lowering total wire length by reducing the use of long wires.
In particular, our cost function assigns a cost of $k-1$ for a net that spans $k$ blocks~\cite{sanchis1993multiple}.

For block floorplanning, we map the $B$ blocks to rectangular areas on the substrate.
Each rectangle's area corresponds to the fraction of total substrate equal to $\rm\frac{components\ in\ block}{total\ components}$, while its shape is flexible.
This step requires that each block after circuit partitioning is large enough so that with component redundancy (i.e., $N^P\approx 2N^L$), the probability that there are enough components in the rectangle area for each block is close to $1$ even in the existence of component deposition imprecision.
The block number $B$ is thus small, so we use simulated annealing (SA) in this step to minimize the estimated inter-block wire length while finding a valid block floorplan.

For circuit design with multiple large top-level modules in RTL description, we pre-partition the circuit by leveraging high-level modularity \textbf{semantics}.
This can lead to better partitioning than the circuit-agnostic FM approach, which produces uniformly sized partitions  and can therefore break natural intra-module locality.
In practice, we partition a large circuit into modules by first leveraging high-level semantics (e.g., RTL modules), floorplan them, further partition them into $B$ blocks in total, and floorplan the $B$ blocks.

Finally, although \pne target rapid in situ manufacturing rather than volume, more than a single instance of the same circuit may be produced in each case. 
The partitioning and floorplanning stage can be reused across circuit instances, while placement and routing are performed from scratch for each circuit instance.
An alternative approach for the latter would be to create a ``template'' solution from the first circuit instance and then adjust it to accommodate deposition-incurred layout differences in the remaining instances. However, the less controlled component deposition is, the less promising such an adjustment-based approach is.
For example, if $C$ component types (pMOS, nMOS, memristor, etc.) are being deposited, and assuming each position on the grid features a given component type with probability $1/C$, deviation from the template at \textit{individual} component granularity becomes the norm rather than the exception.
The coarse-grained blocks produced by partitioning are significantly less sensitive to such variations, hence reusable across circuit instances.

\endgroup

\section{Placement and Routing for \pne}
\label{sec:basic-algos}

We first introduce our formal notations and reiterate the unique aspects of \pne affecting placement and routing in \cref{sec:formalization}.
We then develop our approaches based on classic algorithms that are widely used in traditional VLSI physical design.
In the component placement step (\cref{sec: SA comp assign}), our approach is based on simulated annealing (SA)~\cite{kirkpatrick1983optimization}.
In the wire routing step (\cref{section:routing}), our approach extends Prim's algorithm to support breadth-first search (BFS) or weighted A*\cite{ebendt2009weighted} routing for intra- or inter-block, respectively.

\subsection{Formal Notations}
\label{sec:formalization}

\begingroup
\renewcommand{\arraystretch}{1.35}
\begin{table}[t]
    \centering
    \caption{Notations of inputs and outputs for \pne placement and routing problem.}
    \label{table:notations}
    \begin{footnotesize}
    \begin{tabular}{|c|p{4.4cm}|}
    \hline
    Notation & Description \\
    \hline\hline
    \multirow{1.8}*{$C^L=\{c^l_{i}|i\in[1,N^L]\}$} & The set of logical components: Component $c^l_i$ is either pMOS or nMOS \\
    \hline
    $P^L_i=\{p^l_{i,1}, p^l_{i,2}, p^l_{i,3}\}$ & The set of 3 pins on $c^l_i$ \\
    \hline
    $IO^L=\{io^l_{i}|i\in[1,S]\}$ & The set of I/O pins \\
    \hline
    \multirow{3.3}*{$Nets=\{net_{i}|i\in[1,E]\}$} & The set of nets that comprises the circuit: Net $net_i$ is a set of pins that have the same voltage potential with every pin belonging to exactly one net \\
    \hline
    $D=\{C^L, P^L, IO^L, Nets\}$ & \textbf{Circuit design input} \\
    \hline\hline
    \multirow{3.3}*{$C^P=\{c^p_i|i\in[1,N^P]\}$} & The set of physical components: Component $c^p_i$ is either pMOS or nMOS, is located at $(x^c_i, y^c_i)$ on the substrate, and has an orientation $\theta^c_i$ \\
    \hline
    \multirow{2.5}*{$P^P_i=\{p^p_{i,1}, p^p_{i,2}, p^p_{i,3}\}$} & The set of 3 pins on $c^p_i$: Pin $p^p_{i,j}$'s location $(x^p_{i,j}, y^p_{i,j})$ is inferred from $(x^c_i, y^c_i)$ and $\theta^c_i$ \\
    \hline
    \multirow{1.8}*{$IO^P=\{io^p_i|i\in[1,S']\}$} & The set of available I/O slots on the boundary of the substrate \\
    \hline
    $F=\{C^P, P^P, IO^P\}$ & \textbf{Component layout input}\\
    \hline\hline
    \multirow{2.5}*{$M^C=\{m^c_{i,j}\}$} & The set of mappings from $C^L$ to $C^P$: A mapping $m^c_{i,j}$ in $M^C$ means assigning $c^l_i$ to $c^p_j$ with the same type \\
    \hline
    \multirow{2.5}*{$M^{IO}=\{m^{io}_{i,j}\}$} & The set of mappings from $IO^L$ to $IO^P$: A mapping $m^{io}_{i,j}$ in $M^{IO}$ means assigning $io^l_i$ to $io^p_j$ \\
    \hline
    $M = M^C \cup M^{IO}$ & \textbf{Placement solution output} \\
    \hline\hline
    \multirow{3.3}*{$R = [r_1, r_2, ... , r_R]$} & \textbf{Routing solution output} as a sequence of wires and insulators: $r_i$ is either a straight wire between two endpoints or an insulation point \\
    \hline
    \end{tabular}
    \end{footnotesize}
    
\end{table}
\endgroup

We list our mathematical notations of inputs and outputs for \pne placement and routing in \cref{table:notations}.
The notations align with the example in \cref{fig:formulation}.
Our objective is to minimize the total print time $Pt = \dfrac{\Psi}{\alpha}$, where $\alpha$ represents the wire printing speed and $\Psi$ represents the total wire length in the routing solution $R$.

While our problem formulation is reminiscent of well-known problems in electronics, we devise solutions for the unique set of challenges presented by \pne. We summarize these unique \pne challenges compared to conventional electronics:
\begin{itemize}[noitemsep,topsep=0pt,leftmargin=*]
    \item Typical physical design problems take circuit design $D$ as an input and produce physical layout $F$ and wire routing solution $R$ as outputs. In \pne, $F$ is the pre-deposited component layout, which is an \textit{input} to the problem, necessitating a new type of output, which is a logical-to-physical component mapping $M$. %
    \item Compared to other mapping problems (e.g., FPGA block assignment), component positions $(x^c_i, y^c_j)$ and directions $\theta^c_i$ in \pne are stochastic, making the layout $F$ non-uniform and unique across individual chips.
    \item Component heterogeneity (i.e., every $c^l_i$ or $c^p_i$ is either pMOS or nMOS) requires that logical-to-physical component types must match in $M$.
    \item While \pne share the primary optimization objective of minimizing total wire length $\Psi$, the algorithm runtime becomes an additional first-order concern.
\end{itemize}

\subsection{SA-Based Component Placement}
\label{sec: SA comp assign}

\cref{algo:placement} shows the component placement step based on SA.
The algorithm takes the circuit design $D$ and the component layout $F$ as inputs, and outputs the component placement solution $M$.
At first, an initial temperature $T$ (line 4) and an initial random solution $M$ (line 5) are decided.
We maintain an exponentially decreasing temperature $T$ by multiplying $T$ by a constant $k_T<1$ at the end of every iteration (line 14).
We also maintain a linearly decreasing neighbor distance $d_n$ by taking the logarithm of temperature $T$ (line 7).
In every iteration, we first randomly pick a mapping $m_{i,j}$ from $M$ (line 8), where $m_{i,j}$ means either mapping from logical component $c^l_i$ to physical component $c^p_j$ if $m_{i,j}\in M^C$, or mapping from I/O pin $io^l_i$ to available slot $io^p_j$ on the boundary if $m_{i,j}\in M^{IO}$.
Then, we generate a new solution $M'$ according to the chosen mapping $m_{i,j}$ and the neighbor distance $d_n$ (line 9).

\setlength{\textfloatsep}{3mm}
\begin{algorithm}[t]
\renewcommand\baselinestretch{0.9}\selectfont
\caption{Component placement based on SA.}
\label{algo:placement}
\begin{footnotesize}
\textbf{Input:} circuit design $D$, and component layout $F$\\
\textbf{Output:} component placement solution $M$\\
\SetKwFunction{FMain}{place}
\SetKwProg{Fn}{Function}{:}{}
\Fn{\FMain{$D$, $F$}}{
    $T$ = initial\_temperature()\\
    $M$ = initial\_random\_placement()\\
    \While{$T > \varepsilon$}{
        $d_n$ = $\log T$ \ccc{linearly decreasing neighbor distance}\\
        $m_{i,j}$ = randomly\_pick\_from$(M)$\\
        $M'$ = new\_random\_placement$(M, m_{i,j}, d_n, F)$\\
        $c$ = cost$(M, D, F)$\\
        $c'$ = cost$(M', D, F)$\\
        \If{$c' < c$ \rm\textbf{or} ${\rm rand}(0, 1) < e^{(c-c')/T}$}{
            $M$ = $M'$
        }
        $T = T*k_T$ \ccc{exponentially decreasing temperature}
    }
    \Return $M$
}
\end{footnotesize}
\end{algorithm}

In detail, assume $m_{i,j}\in M^C$, meaning $m_{i,j}$ is $m^c_{i,j}$.
We first randomly pick another physical component $y$ of the same type within distance $d_n$ from $j$.
If $y$ is not yet mapped to any other logical component (i.e., $\forall x, m^c_{x,y}\notin M^C$), we simply remove $m^c_{i,j}$ and add $m^c_{i,y}$ to $M^C$.
If $y$ is mapped to another logical component $x$ (i.e., $m^c_{x,y}\in M^C$), we swap the mapping (i.e., replace $m^c_{i,j}$ and $m^c_{x,y}$ with $m^c_{i,y}$ and $m^c_{x,j}$ in $M^C$).
This rule also applies if $m_{i,j}\in M^{IO}$.
The cost of a component placement solution $M$ (and $M'$) is obtained by calculating the length of a minimum spanning tree (MST) on every net and taking the summation (lines 10--11).
This is because the MST length is highly correlated to the routed total wire length $\Psi$~\cite{huang1997partitioning}.
Our own evaluations (\cref{sec:eval_breakdowns}) additionally show that reducing total wire length also results in fewer insulation points $\Omega$.
If the new cost $c'$ is less than $c$, or with a probability of $e^{(c-c')/T}$, we update $M$ with $M'$ (lines 12--13).

However, such an approach is extremely slow, taking one week to run on a circuit with only $\sim$5,000 logical components.
We therefore add optimizations to improve its performance.
In every iteration, at most two mapping elements are modified, so only a small number (i.e., $\leq$ 6) of net costs must be recalculated.
This constraint reduces redundant MST recalculation, whose complexity is $O(K^2)$ for a net with $K$ pins.
In addition, calculating MST for large nets in every iteration is very slow.
These nets are usually power supplies and clocks, which span most of the circuit and are therefore not amenable to localization.
Given that only $<0.5\%$ of nets in real circuits are large (pin count $\geq$ 32)~\cite{chen2019salt, yang2023towards}, we opt to ignore them when calculating the cost at each iteration.

Component placement is performed for each block independently. 
However, in a net spanning multiple blocks (i.e., inter-block net), components in a given block will be connected to components in other blocks.
Thus, we devise a heuristic that maps components in an inter-block net closer to each other.
If one $net$ spans both blocks A and B, we want $net$'s pins in block A as close to block B as possible and vice versa.
To introduce that bias in component placement, we introduce \textbf{virtual pins}.
These virtual pins are not used in routing but only to improve the quality of placement.
For example, when performing component placement in block A, we only consider the subset $net_A$ of $net$.
We add a virtual pin to $net_A$ at the center of block B, and vice versa when performing component placement in block B.
SA will thus automatically be biased to map components in $net$ close to the boundary of blocks A and B.

When a single $net$ spans more than 2 blocks, we extend our heuristic to calculate the minimum spanning tree (MST) of these blocks.
When performing component placement in block $i$, we consider the subnet $net_i$.
We add a virtual pin to $net_i$ at the center of block $j$ such that $j$ is $i$'s parent in the MST.
If block $i$ is the root of the MST, we select $j$ such that it is $i$'s nearest child.
The intuition behind our heuristic is to introduce bias in the placement decision made by SA while avoiding increasing the algorithm's runtime.
Adding more virtual pins (e.g., at the center of all neighbor blocks in the MST) can improve placement quality.
However, it will introduce a higher overhead in calculating the SA cost function.
Adding a single virtual pin increases the algorithm's runtime by 6\% (\cref{sec:eval_breakdowns}); thus, we limit the number of virtual pins we consider to one.

\subsection{Prim's and BFS-/Weighted A*-Based Wire Routing}
\label{section:routing}

\begin{figure}[t]
    \centering
    \subfloat[Euclidean space.] {\label{fig:euclidean_overlap}\includegraphics[width=0.48\columnwidth]{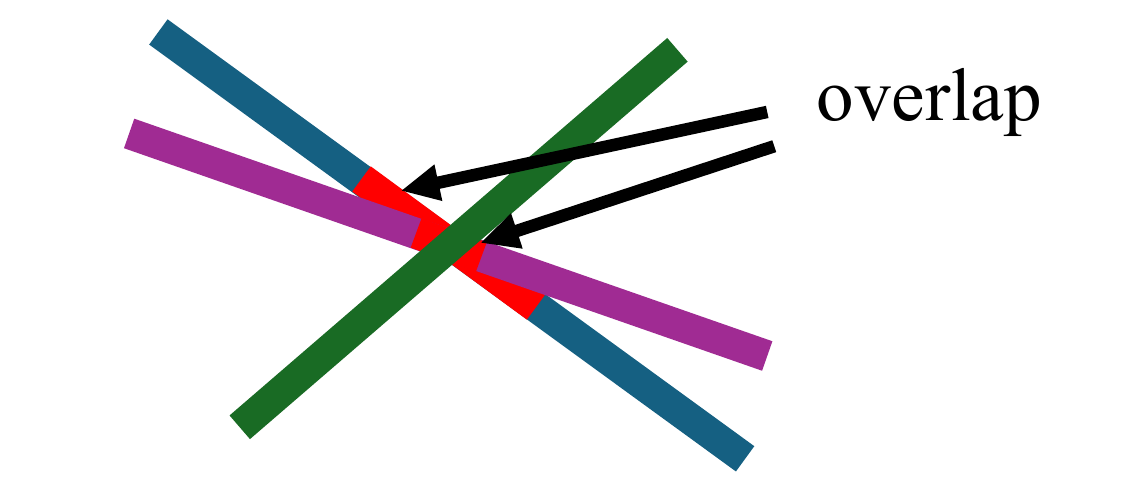}}
    \subfloat[Manhattan space.] {\label{fig:manhattan_overlap}\includegraphics[width=0.48\columnwidth]{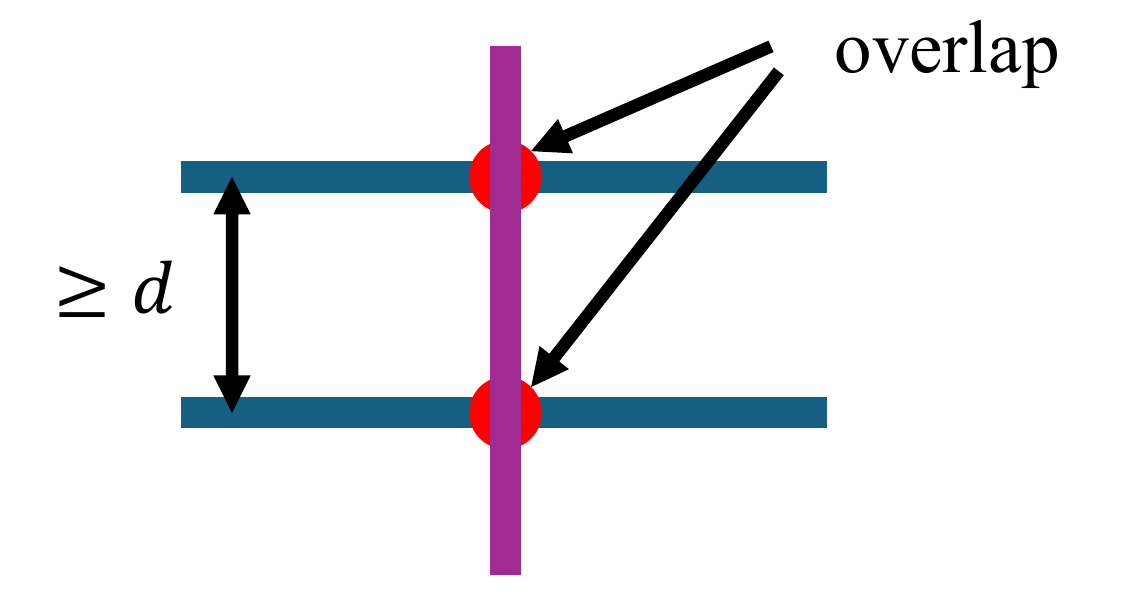}}
    \caption{Wire overlaps in Euclidean and Manhattan space.}
    \label{fig:overlap}
\end{figure}

\setlength{\textfloatsep}{3mm}
\begin{algorithm}[b]
\renewcommand\baselinestretch{0.9}\selectfont

\caption{Wire routing based on Prim's and BFS.}
\label{algo:routing}
\begin{footnotesize}
\textbf{Input:} circuit design $D$, and component layout $F$ and placement $M$ \\
\textbf{Output:} wire routing solution $R$\ccc{wire and insulator sequence}\\
\textbf{Parameter:} grid length $g$\\
\SetKwFunction{FMain}{route}
\SetKwProg{Fn}{Function}{:}{}
\Fn{\FMain{}}{
    $R$ = $[]$; $G$ = create\_discretized\_graph($F$, $g$)\\
    $G$ = remove\_component\_bounding\_boxes($G$, $M$)\\
    $O^{net}$ = decide\_net\_order($D$, $M$, $G$)\\
    \For{$net\in O^{net}$}{
        $O^{pin}$ = pin\_order\_by\_Prims($net$)\\
        $G$ = reconnect($G$, $O^{pin}_1$); $S^{t}$ = $\{O^{pin}_1\}$\\
        $\{R_{insu}, R_{wire}\}$ = $\{\}$\\
        \For{$pin\in O^{pin} - [O^{pin}_1]$}{
            $G$ = reconnect($G$, $pin$)\\
            $\{R_{insu}, R_{wire}\}$ += BFS($s=pin$, $t=S^{t}$, $G$)\\
            $S^{t}$ = vertices\_on($R_{wire}$)
        }
        $G$ = remove\_edges\_and\_mark\_vertices($G$, $R_{wire}$)\\
        $R$.push\_back($R_{insu}$); $R$.push\_back($R_{wire}$)
    }
    \Return $R$
}
\end{footnotesize}
\end{algorithm}

Wire-printing nozzles can travel freely in Euclidean space on the 2D substrate, hence wires are not constrained to following horizontal or vertical channels like in traditional VLSI physical design.
Existing works either discretize the space by triangular Delaunay triangulation~\cite{chung2023any, chen2022obstacle} or immediately calculate the minimum path by quasi-Newton methods~\cite{kohira2012any, honda2014acceleration}.
These approaches are slow, only fitting circuits with hundreds of nets.
However, while an option worth considering, we find that routing in Euclidean space also introduces additional difficulties in printing compared to the alternative conventional option of routing in Manhattan space.
\cref{fig:overlap} shows the comparison of wire overlaps in Euclidean and Manhattan space.
Red areas show the overlaps.
In Euclidean space, the overlap area of two wires is dictated by their intersection angle, raising the following problems:
\begin{itemize}[noitemsep,topsep=0pt,leftmargin=*]
    \item Sharp angles between crossing wires result in long-running overlaps, increasing concerns about electromagnetic interference.
    \item Long overlap requires printing several adjacent insulation points, potentially challenging our current assumption that insulation printing time can be hidden by wire printing time.
    \item If multiple wires and insulators are printed at the same position, the additive process of multiple wires and insulators can result in a considerable localized increase in the circuit's vertical dimension, potentially introducing mechanical concerns.
\end{itemize}
In contrast, in Manhattan space, if two wires run in parallel, we just need to maintain a minimum distance $d$ between them.
If two wires cross each other, the overlap that must be insulated is always a point.
Further, imposing such a limitation prevents having more than two wires overlap at the same point.

For Manhattan routing, we discretize the substrate into a grid. %
The spacing between adjacent grid points is  $g\geq d$.
\cref{algo:routing} describes the routing step based on Prim's algorithm and BFS, taking the logical circuit design $D$, the physical component layout $F$, and the component placement outputs $M$ of the previous step as inputs.
It outputs the wire routing solution $R$ as a wire and insulator sequence.
Initially, a discretized graph $G$ represents the grid (line 5):
grid points are vertices, and grid connections are $G$'s edges.
Each edge's weight is set to $1$.
We simplify the printing process by avoiding insulations and wires crossing over components, as follows.
Every mapped physical component is encased in a \textit{bounding box} and edges inside bounding boxes are removed from $G$ forever (line 6).
The edges connected to the extensions of pins outside the bounding box are temporarily removed (line 6) and reconnected when the pin is being routed (lines 10, 13), preventing the vertex from being occupied by other nets.

After the initialization of the graph, we search for the wire routing outputs net by net.
We first decide an order $O^{net}$ of nets by sorting the number of pins in every net's \textit{net bounding box} (line 7)~\cite{kahng2011vlsi}.
Then, to search the routing outputs of every $net\in O^{net}$, we extend Prim's algorithm.
We refer to Prim's pin processing order $O^{pin}$ (line 9) and replace the edge connecting step in Prim's algorithm by BFS to find valid paths in the discretized graph (line 14).
The target vertex set ($S^{t}$) in BFS consists of the vertices on routed wires $R_{wire}$ in $net$ (line 16) except for the first iteration, where the target vertex is $O^{pin}_1$ (line 10).
The routed solution for a net is a Steiner tree~\cite{hwang1992steiner}, which gives a shorter total wire length than a spanning tree.
Finally, after routing for every net, we remove edges and mark vertices as visited on the routed path (line 16).
The removal of edges prevents complete wire overlap, while vertex marking facilitates identifying wire intersections.
We add results to the output after each net's routing is completed by first adding insulators and then wires (line 17).
The final output $R$ contains the information of both positions and print order.

Inter-block routing can also be performed using \cref{algo:routing}.
However, the graph size of inter-block routing is much larger than the graph size of the individual blocks.
Thus, instead of using BFS that produces optimal results with expected complexity $O(\phi^2)$, where $\phi$ is the grid Manhattan distance between the vertices connected, we use A*~\cite{hart1968formal} with negligible optimality sacrifice (shown in \cref{sec:eval_main}).
Specifically, we use \textbf{weighted A*}~\cite{ebendt2009weighted}, whose expected complexity is $O(\phi)$ in our scenario because wires do not have to take long detours when reaching between two vertices.
A* requires specifying both the starting and the target vertex (i.e., it cannot efficiently connect to the closest vertex in a set of vertices).
Thus, we find the closest pair of pins between the start net and the target net, and then use A* to connect them.
We use reaching any vertex on the target net as a termination condition for the algorithm.

\section{Evaluation}
\label{sec:eval}

We implement our algorithms in C++ and evaluate their performance on an AMD EPYC 9754 processor. %
\cref{table:dataset} shows the five circuit designs of increasing size used in our experiments.
We generate each gate-level circuit from Verilog code using Synopsys Design Compiler~\cite{compiler2016synopsys} and the transistor-level circuit by providing a %
conversion of each gate type to its transistor-level implementation.
Throughout \cref{sec:eval}, a ``component'' corresponds to a transistor.

The selected circuits are representative of the target application scenarios of \pne, while the circuit sizes we evaluate span a wide range. 
\pne's biggest advantage of drastically lower capital costs than traditional electronics makes the technology suitable for desktop chip printing and rapid iteration, instead of mass production.
Hence, it is a domain-specific technology rather than a general-purpose replacement for traditional manufacturing methods.
For example, one application scenario is printing a custom simple chip that collects and preprocesses data from outdoor sensors.
Such a chip could, for example, contain multiple perceptrons, corresponding to Cir.2.
Another application scenario is developing a dedicated module or mini-core, which requires low cost and fast prototyping.
This corresponds to Cir.3--Cir.5.
It is unlikely that \pne will be used to manufacture chips with hundreds of millions of transistors or more, such as a high-performance CPU or GPU.
Therefore, we limit our benchmark circuit size to the sub-million component range.

For a circuit with $N^L$ components, we randomly generate the approximate mesh component layout with $N^P = 2N^L$ components in a $W\times H$ area with $W=H=\sqrt{N^P}\times \rho$, where $\rho$ is the average inter-component distance.
We use values representative of the current state of our target \pne technology~\cite{filler2023nanomodular}:
pMOS/nMOS transistor length of $150/130\ nm$, respectively, wire width $w=100\ nm$, and
$\rho=10\ \mu m$.
We report runtime $Rt$, total wire length $\Psi$, and insulator number $\Omega$. %
Because $\Psi$ directly depends on $\rho$, we report relative total wire length $\Psi_r = \Psi / \rho$ instead of the absolute $\Psi$ value.
A denser technology with $\rho' < \rho$ will result in unchanged $\Psi_r$, and, equivalently, in a reduction of absolute wire length to $\Psi\times \rho'/\rho$.

\begin{table}[t]
    \centering
    \caption{Evaluated circuits. $N^L$: component number. $S$: I/O pin number. $E$: net number.}
    \label{table:dataset}
    \begin{footnotesize}
    \begin{tabular}{|c|c|c|c|c|}
    \hline
    Circuit & Description & $N^L$ & $S$ & $E$ \\
    \hline\hline
    Cir.1 & 1-bit full adder & 42 & 7 & 28 \\
    \hline
    \multirow{2}*{Cir.2} & Perceptron & \multirow{2}*{4,976} & \multirow{2}*{21} & \multirow{2}*{2,508}\\
     & with four 4-bit inputs & & & \\
    \hline
    Cir.3 & 32-bit ALU & 39,086 & 103 & 19,613\\
    \hline
    \multirow{2}*{Cir.4} & 32-bit RISC-V core & \multirow{2}*{147,522} & \multirow{2}*{69} & \multirow{2}*{73,797}\\
     & single cycle~\cite{riscvsingle} & & & \\
    \hline
    \multirow{2}*{Cir.5} & 32-bit RISC-V core & \multirow{2}*{350,884} & \multirow{2}*{269} & \multirow{2}*{175,592}\\
     & 5-stage pipelined~\cite{riscvpipeline} & & & \\
    \hline
    \end{tabular}
    \end{footnotesize}
\end{table}

\begin{table*}[t]
    \centering
    \caption{Comparisons of different approaches on dataset. %
    $\Psi_r$: relative total wire length. $\Omega$: number of insulation points. 
    $Rt$: Runtime. For the last three approaches, $Rt$ is reported as \textit{X+Y}, where \textit{X} refers to the partitioning and floorplaning stage, while \textit{Y} refers to the placement and routing stage. When printing multiple instances of the same circuit, \textit{X} is incurred only once, while \textit{Y} is incurred per circuit instance.
    We omit bottom-left results (circuits too small to apply partitioning) and top-right results ($Rt$ exceeds a day---e.g., see $Rt$ for \CodeIn{Small} on Cir.4).}
    \label{table:overall_performance}
    \begin{footnotesize}
    \begin{tabular}{|c|c|c|c|c|c|c|c|c|c|c|c|}
    \hline
    \multirow{2}*{Approach} & \multirow{2}*{Metric} & \multicolumn{2}{c|}{Cir.1} & \multicolumn{2}{c|}{Cir.2} & \multicolumn{2}{c|}{Cir.3} & \multicolumn{2}{c|}{Cir.4} & \multicolumn{2}{c|}{Cir.5}\\
    \cline{3-12}
     & & Absolute & Relative & Absolute & Relative & Absolute & Relative & Absolute & Relative & Absolute & Relative\\
    \hline\hline
    
    \multirowcell{2}{\CodeIn{SA in}\\\CodeIn{Euclidean}}
     & $\Psi_r$ & 123 & 0.732 & 25.6K & 0.734 & 301K & 0.788 & - & - & - & -\\
     & $\Omega$ & 1,132 & 9.433 & 211K & 5.553 & 2.19M & 3.967 & - & - & - & -\\
    
    \hline\hline

    \multirowcell{2}{\CodeIn{SA in}\\\CodeIn{Manhattan}}
     & $\Psi_r$ & 144 & 0.857 & 31.7K & 0.908 & 371K & 0.971 & - & - & - & -\\
     & $\Omega$ & 2,184 & 18.200 & 302K & 7.947 & 2.49M & 4.511 & - & - & - & -\\
    
    \hline\hline

    \multirow{3}*{\CodeIn{Small}} & $Rt$ & $<$0.1 s & - & 147 s & 1.000 & 4.2 hrs & 1.000 & 2.8 days & 1.000 & - & -\\
     & $\Psi_r$ & 168 & 1.000 & 34.9K & 1.000 & 382K & 1.000 & 1.64M & 1.000 & - & -\\
     & $\Omega$ & 120 & 1.000 & 38K & 1.000 & 552K & 1.000 & 2.6M & 1.000 & - & -\\
    
    \hline\hline

    \multirow{3}*{\CodeIn{Large\&BFS}} & $Rt$ & - & - & 1+24 s & 0.170 & 3+353 s & 0.024 & 35+2680 s & 0.011 & - & -\\
     & $\Psi_r$ & - & - & 36.4K & 1.043 & 582K & 1.524 & 2.89M & 1.762 & - & -\\
     & $\Omega$ & - & - & 43K & 1.132 & 1.45M & 2.627 & 9.1M & 3.500 & - & -\\
    
    \hline\hline

    \multirow{3}*{\CodeIn{Large\&A*}} & $Rt$ & - & - & 1+9 s & 0.068 & 3+165 s & 0.011 & 35+217 s & 0.001 & 235+1142 s & 1.000\\
     & $\Psi_r$ & - & - & 36.6K & 1.049 & 591K & 1.547 & 2.95M & 1.799 & 13.9M & 1.000\\
     & $\Omega$ & - & - & 45K & 1.184 & 1.48M & 2.681 & 9.4M & 3.615 & 91.2M & 1.000\\
    
    \hline\hline

    \multirowcell{3}{\CodeIn{Large\&A*}\\\CodeIn{+}\\\CodeIn{Semantics}} & $Rt$ & - & - & - & - & 2+195 s & 0.013 & 17+237 s & 0.001 & 48+819 s & 0.630\\
     & $\Psi_r$ & - & - & - & - & 464K & 1.215 & 1.98M & 1.207 & 6.9M & 0.496\\
     & $\Omega$ & - & - & - & - & 873K & 1.582 & 4.0M & 1.538 & 19.6M & 0.215\\
    
    \hline
    \end{tabular}
    \end{footnotesize}

\end{table*}

We evaluate the six approaches listed in \cref{table:overall_performance}'s first column.
The first two denote direct wire connection by MST (i.e., without BFS routing) after component placement by simulated annealing (SA) in Euclidean and Manhattan space, respectively.
We assume that an insulator covers a rectangular area $w\times w$.
Therefore, in these two approaches, we use $\lceil l/w \rceil$ insulators to cover overlaps with length $l > w$.
\CodeIn{Small} represents the bypassed workflow for small circuits shown in \cref{fig:new_workflow}.
\CodeIn{Large\&BFS} denotes the full workflow for large circuits with the virtual pin optimization and BFS for inter-block routing.
\CodeIn{Large\&A*} additionally applies the weighted A* algorithm optimization.
Finally, \CodeIn{Large\&A*+Semantics} also incorporates high-level semantics derived from component declarations in Verilog.
We configure the iteration number $T_N$ of SA (inferred by $T$, $k_T$ and $\varepsilon$ from \cref{algo:placement}) for a block with $N$ components to $T_N = N^2$.
The partition number used for Cir.2--Cir.5 is 4, 9, 30, 64, respectively.

\subsection{Main Results}
\label{sec:eval_main}

\cref{table:overall_performance} compares the performance of the six approaches, reporting both absolute and relative values.
Relative values are normalized to \CodeIn{Small} as baseline for Cir.1--Cir.4, and to \CodeIn{Large\&A*} for Cir.5.

We first focus on the effect of the first three approaches on $\Psi_r$ and $\Omega$.
\CodeIn{SA in Euclidean} produces the shortest attainable $\Psi_r$, but it generates poor routing results because of sharp wire angles and long overlaps, as discussed in \cref{section:routing}.
We therefore use it only as a lower-bound target for $\Psi_r$.
On Cir.1--Cir.3, we find that compared to direct connection in Euclidean space, \CodeIn{Small} reduces $\Omega$ by $6.3\times$, at the cost of $25.2\%$ higher $\Psi_r$.
Compared to direct connection in Manhattan space, \CodeIn{Small} reduces $\Omega$ by $10.2\times$ with $8.8\%$ higher $\Psi_r$.

\textit{The last four approaches achieve huge speedup in $Rt$ (i.e., time-to-solution) by modestly sacrificing $\Psi_r$.}
The tradeoff between $Rt$, $\Psi_r$, and $\Omega$ varies in Cir.2--Cir.4 by circuit scale and partition number.
For small circuit Cir.2, circuit partitioning achieves $5.9\times$ speedup with only $4.3\%$ increase in $\Psi_r$ and $13.2\%$ increase in $\Omega$.
We further achieve $2.5\times$ speedup by A* with negligible increase ($0.5\%$) in $\Psi_r$.
The total tradeoff on Cir.2 is $14.7\times$ speedup with $4.9\%$ increase in $\Psi_r$ and $18.4\%$ increase in $\Omega$.
For large circuit Cir.4, the tradeoff is even more valuable, achieving a $952\times$ speedup for only a $21\%$ increase in $\Psi_r$ and $54\%$ increase in $\Omega$.

By comparing the last two approaches, we find circuit partitioning by semantics effective in reducing $\Psi_r$.
We find $34.9\%$ and $59.0\%$ average decrease in $\Psi_r$ and $\Omega$ on Cir.3--Cir.5.
The $Rt$ increase is $17.3\%$ on Cir.3 and only $0.8\%$ on Cir.4.
Leveraging semantics even achieves a $1.6\times$ speedup on Cir.5 because of the largely decreased $Rt$ for inter-block routing.
It also reduces runtime for partitioning and floorplaning stage by splitting the stage into two levels.
We further investigate partitioning by semantics in \cref{sec:eval_breakdowns}.

\subsection{End-to-End Analysis}

\begin{table}[t]
    \centering
    \caption{Comparison of algorithm runtime, circuit printing time, and end-to-end manufacturing time of one Cir.4 instance on different approaches with $1\ mm/s$ (top) and $1\ cm/s$ (bottom) wire printing speed. Speedup is relative to the \CodeIn{Small} approach.}
    \label{table:end-to-end}
    \begin{footnotesize}
    \begin{tabular}{|c|c|c|c|c|}
    \hline
    Approach & Algorithm & Printing & End-to-end & Speedup \\
    \hline\hline
    \CodeIn{Small} & 2.8 days & 273.33 min & 3.0 days & $1.0\times$ \\
    \hline
    \CodeIn{Large\&BFS} & 45.25 min & 481.67 min & 526.92 min & $8.2\times$ \\
    \hline
    \CodeIn{Large\&A*} & 4.20 min & 491.67 min & 495.87 min & $8.7\times$ \\
    \hline
    \CodeIn{Large\&A*+} & \multirow{2}*{4.23 min} & \multirow{2}*{330.00 min} & \multirow{2}*{334.23 min} & \multirow{2}*{$12.9\times$} \\
    \CodeIn{Semantics} & & & & \\
    \hline
    \multicolumn{5}{c}{}\\
    \hline
    \CodeIn{Small} & 2.8 days & 27.33 min & 2.8 days & $1.0\times$ \\
    \hline
    \CodeIn{Large\&BFS} & 45.25 min & 48.17 min & 93.42 min & $43.2\times$ \\
    \hline
    \CodeIn{Large\&A*} & 4.20 min & 49.17 min & 53.37 min & $75.5\times$ \\
    \hline
    \CodeIn{Large\&A*+} & \multirow{2}*{4.23 min} & \multirow{2}*{33.00 min} & \multirow{2}*{37.23 min} & \multirow{2}*{$108.3\times$} \\
    \CodeIn{Semantics} & & & & \\
    \hline
    \end{tabular}
    \end{footnotesize}
    \vspace{2mm}
\end{table}

In this section, we illustrate the joint contribution of the improved performance in algorithm runtime and increased circuit printing time from an end-to-end manufacturing standpoint.
Consider two NE technology instances with average inter-component distance of $10\ \mu m$ and different wire printing speeds ($\alpha$): $1\ mm/s$ (current \pne state) and $1\ cm/s$ (near-future expectation), respectively.
We translate algorithm runtime and wire length for \cref{table:overall_performance}'s  Cir.4 to end-to-end manufacturing time ($Et$) for one circuit instance.
$Et$ is calculated by adding algorithm runtime $Rt$ and circuit printing time $Pt=\dfrac{\Psi_r\times10\ \mu m}{\alpha}$.
\cref{table:end-to-end} shows the end-to-end manufacturing times and speedup over the \CodeIn{Small} approach.

The \CodeIn{Large\&A*+Semantics} approach's higher wire length increases printing time from 27 to 33 minutes. 
However, \CodeIn{Large\&A*+ Semantics} produces this slightly suboptimal solution $952\times$ faster (4.23 minutes) than the \CodeIn{Small} approach (2.8 days). Overall, considering end-to-end manufacturing time, the tradeoff is very clear: despite the longer wire length, total manufacturing time is $108.3\times$ faster. Even with $10\times$ slower wire printing speed (i.e., $1\ mm/s$), \CodeIn{Large\&A*+Semantics} still presents a favorable tradeoff, resulting in $12.9\times$ lower end-to-end manufacturing time.

The results highlight the balance required between manufacturing time and circuit performance.
Our proposed algorithms achieve substantial speedup in end-to-end manufacturing time, allowing the manufacturing of individual chips in minutes instead of days. However, this speedup comes at the expense of increasing the circuit's wire length by 21\%, potentially degrading the circuit's frequency.

\begin{figure}[t]
    \centering
    \subfloat[Runtime ($Rt$).] {\label{fig:evaluations/perceptron_runtime}\includegraphics[width=\columnwidth]{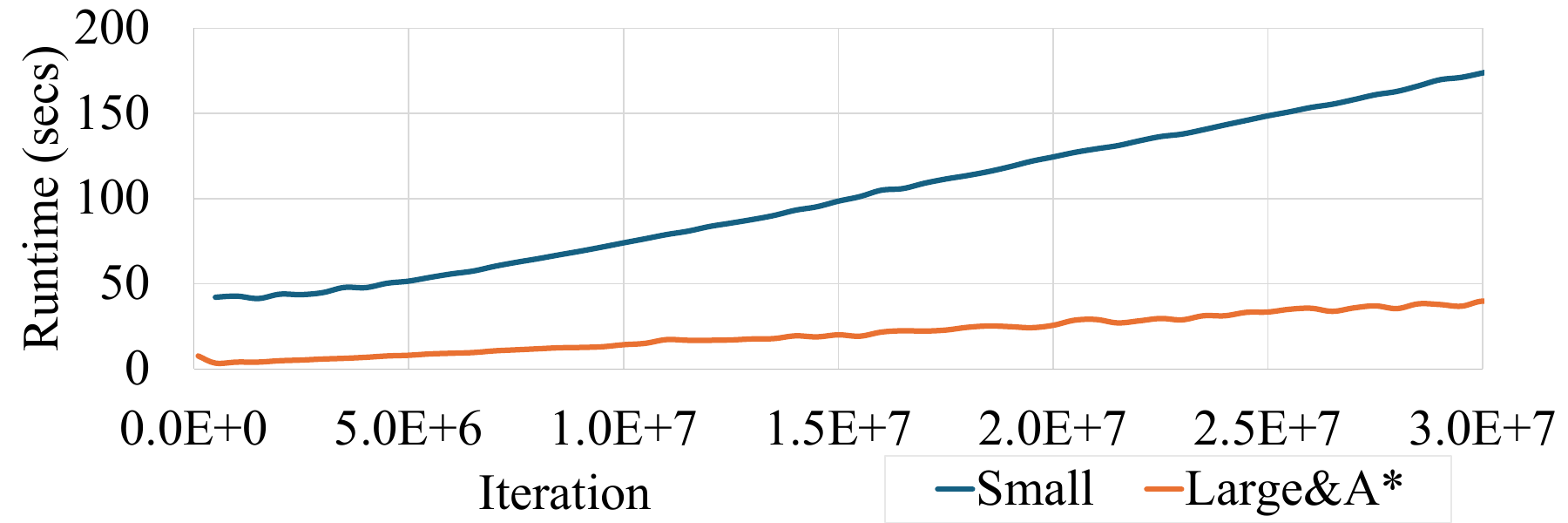}}\\
    \subfloat[Relative total wire length ($\Psi_r$).]{\label{fig:evaluations/perceptron_twl}\includegraphics[width=\columnwidth]{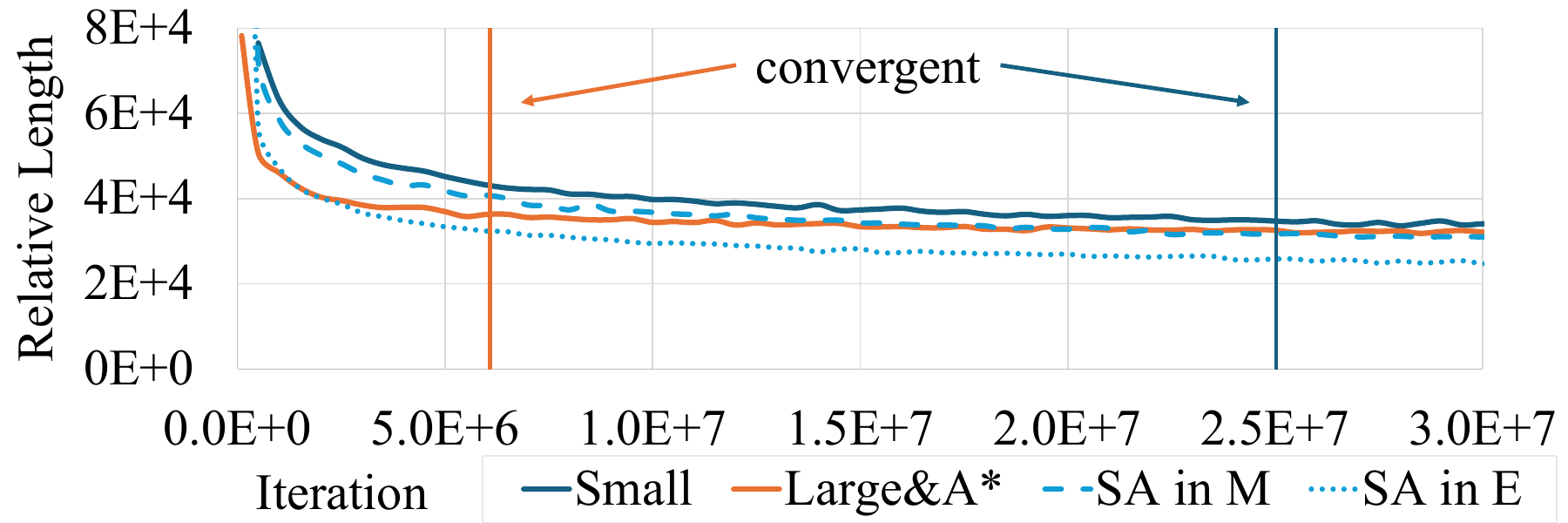}}\\
    \subfloat[Number of insulation points ($\Omega$).] {\label{fig:evaluations/perceptron_insu}\includegraphics[width=\columnwidth]{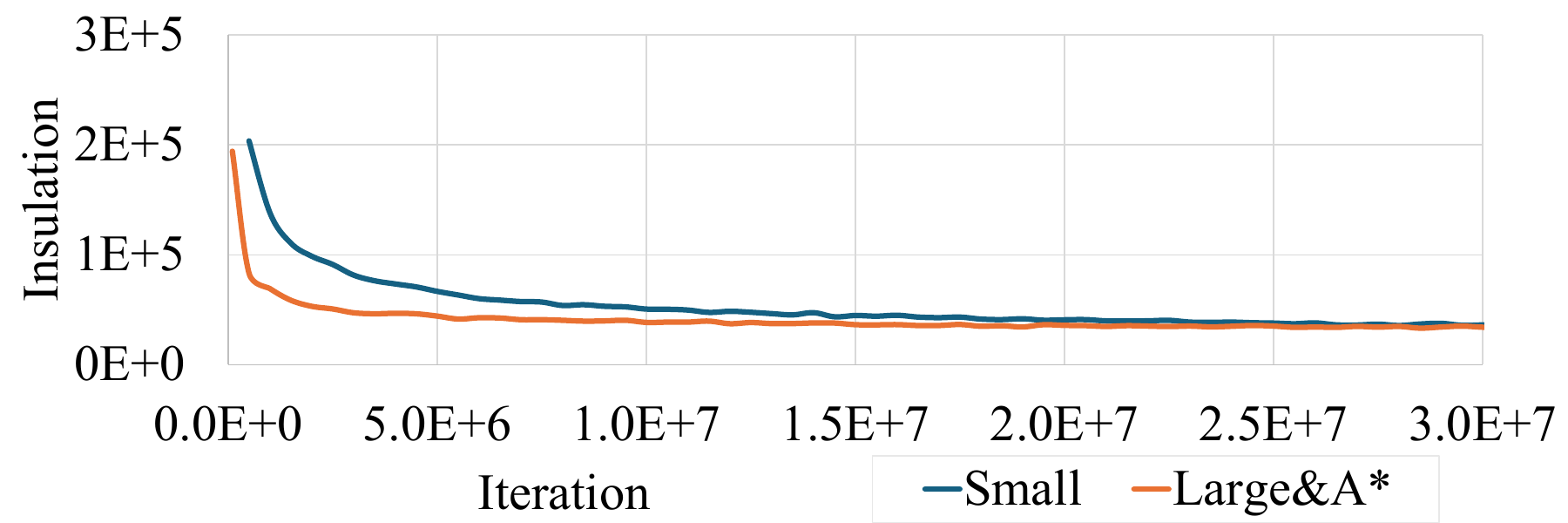}}
    \caption{Effect of algorithm iteration number on Cir.2.}
    \label{fig:evaluations_perceptron}
    \bigskip
\end{figure}

\begin{figure}[t]
    \centering
    \subfloat[Runtime ($Rt$).] {\label{fig:evaluations/riscv_runtime}\includegraphics[width=\columnwidth]{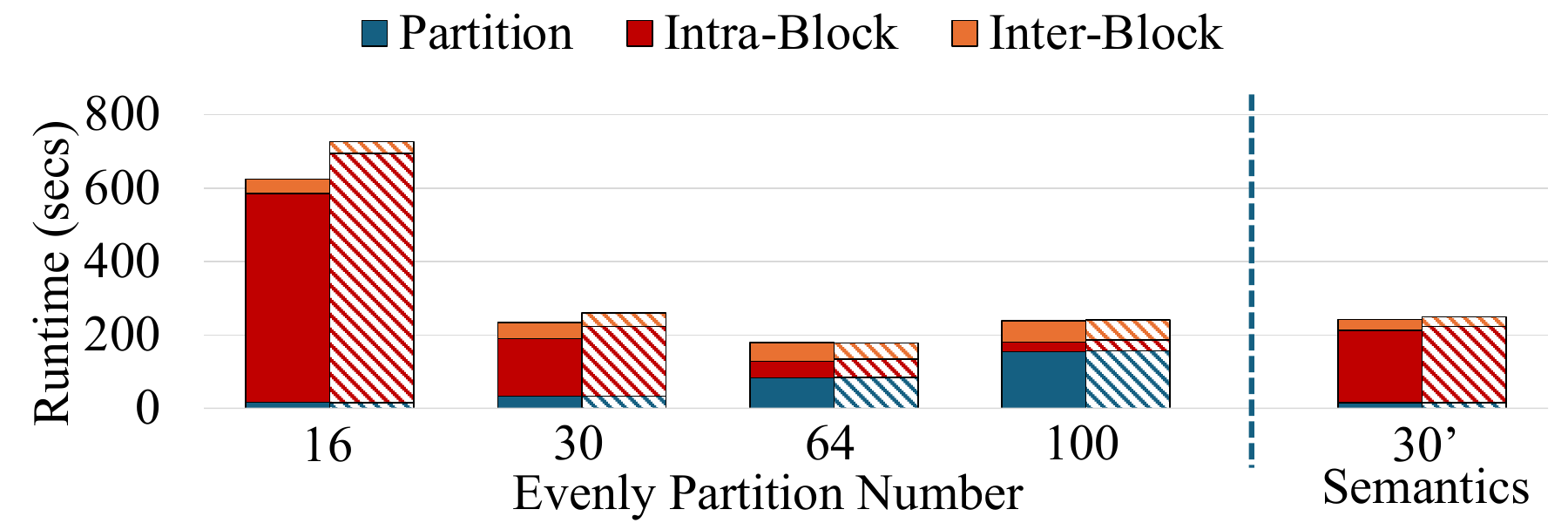}}\\
    \subfloat[Relative total wire length ($\Psi_r$).] {\label{fig:evaluations/riscv_twl}\includegraphics[width=\columnwidth]{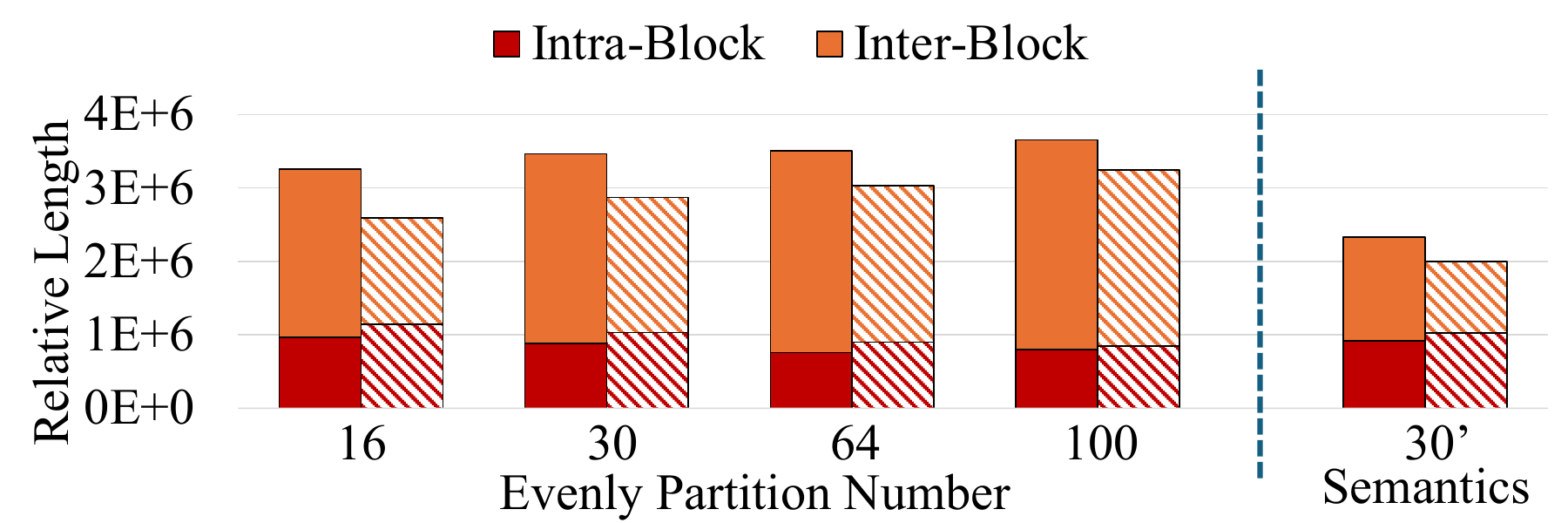}}\\
    \subfloat[Number of insulation points ($\Omega$).] {\label{fig:evaluations/riscv_insu}\includegraphics[width=\columnwidth]{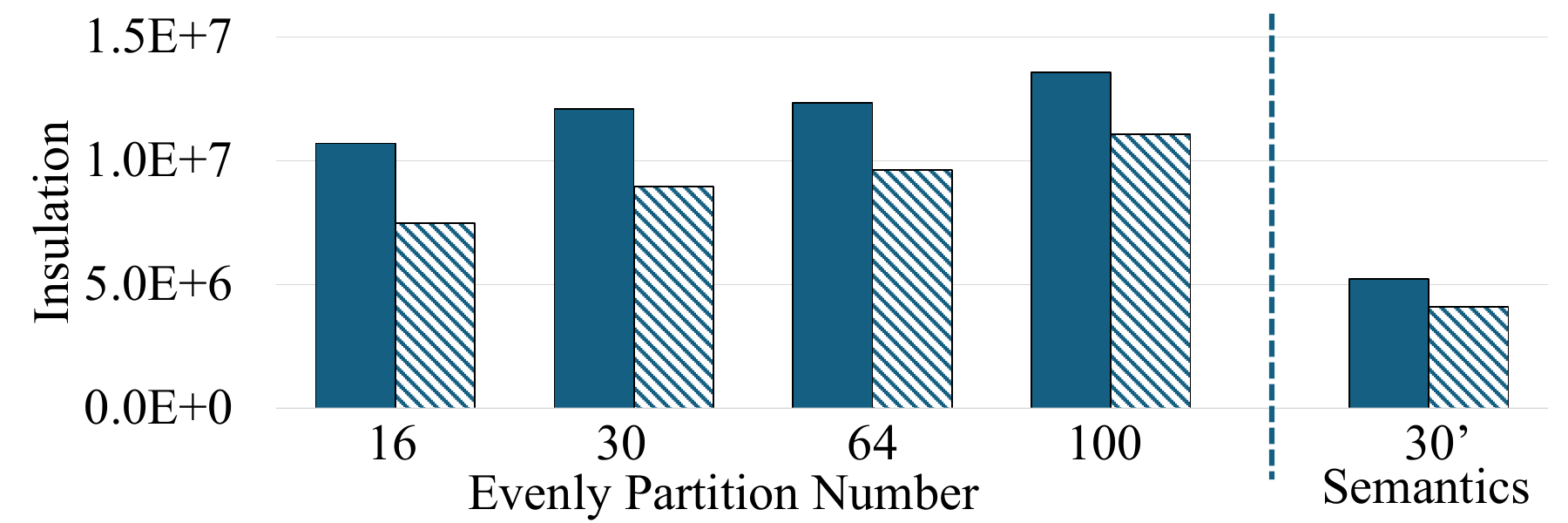}}
    \caption{Effect of partition number, virtual pin, and semantics on Cir.4. Solid/Shaded bars: virtual pin off/on.}
    \label{fig:evaluations_riscv}
\end{figure}

\begin{figure*}[t]
    \centering
    \subfloat[Cir.2 without partitioning.] {\label{fig:visualization/perceptron_single}\includegraphics[width=0.33\textwidth]{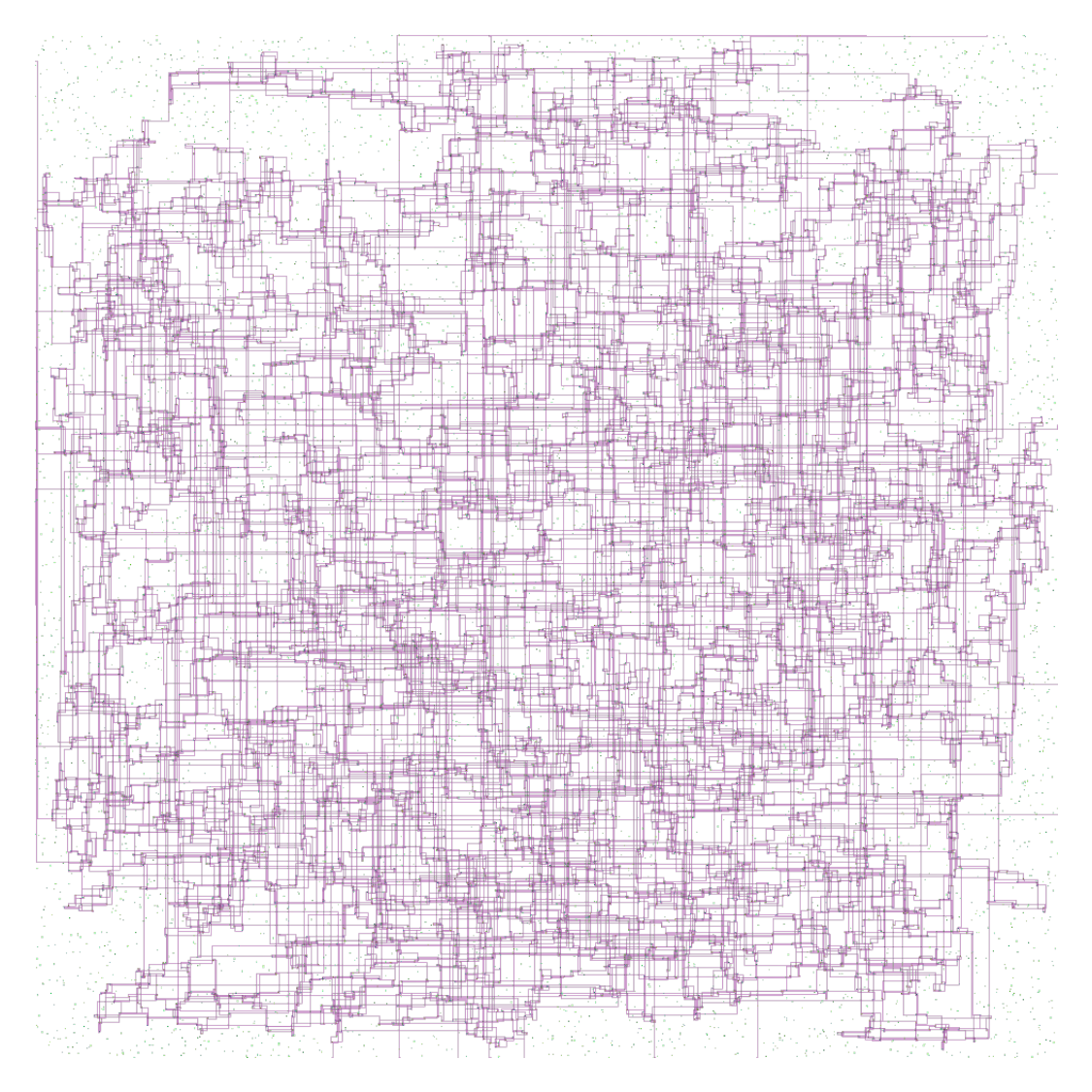}}
    \subfloat[Detailed view of sub-area in (a).] {\label{fig:visualization/perceptron_detail}\includegraphics[width=0.33\textwidth]{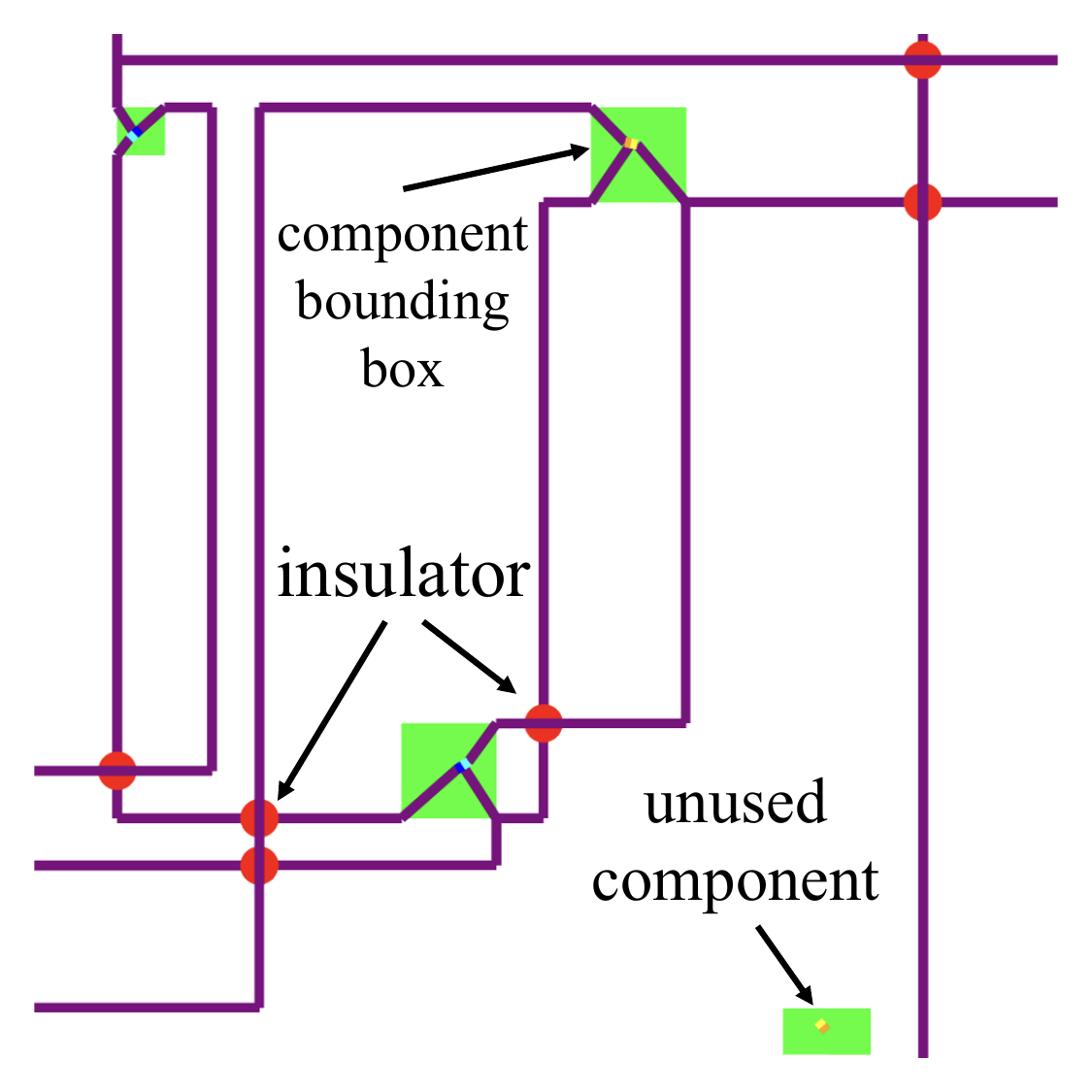}}
    \subfloat[Cir.2 with four evenly-sized partitions.] {\label{fig:visualization/perceptron_parallel}\includegraphics[width=0.33\textwidth]{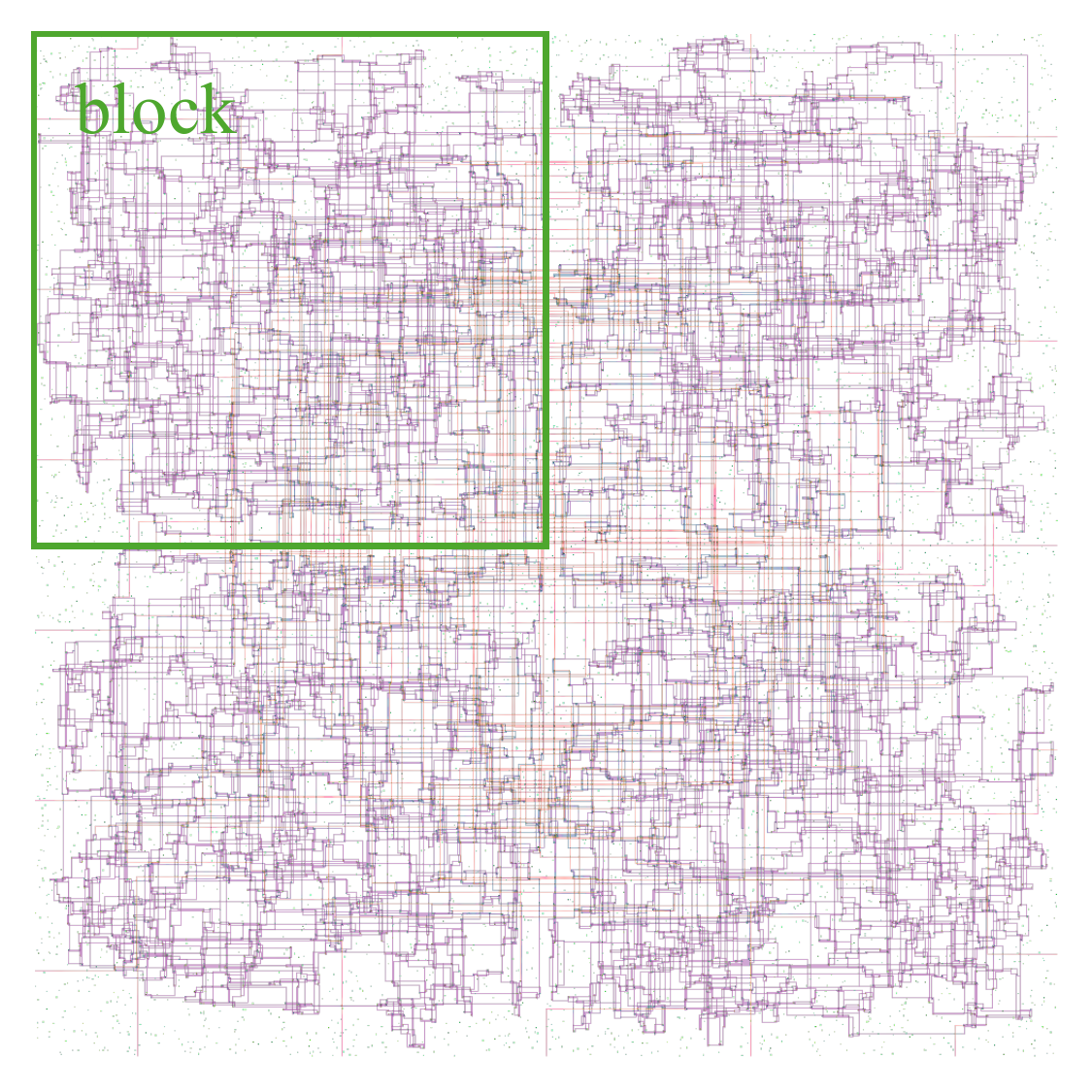}}\\
    \subfloat[Cir.4 without partitioning.] {\label{fig:visualization/riscv_single}\includegraphics[width=0.33\textwidth]{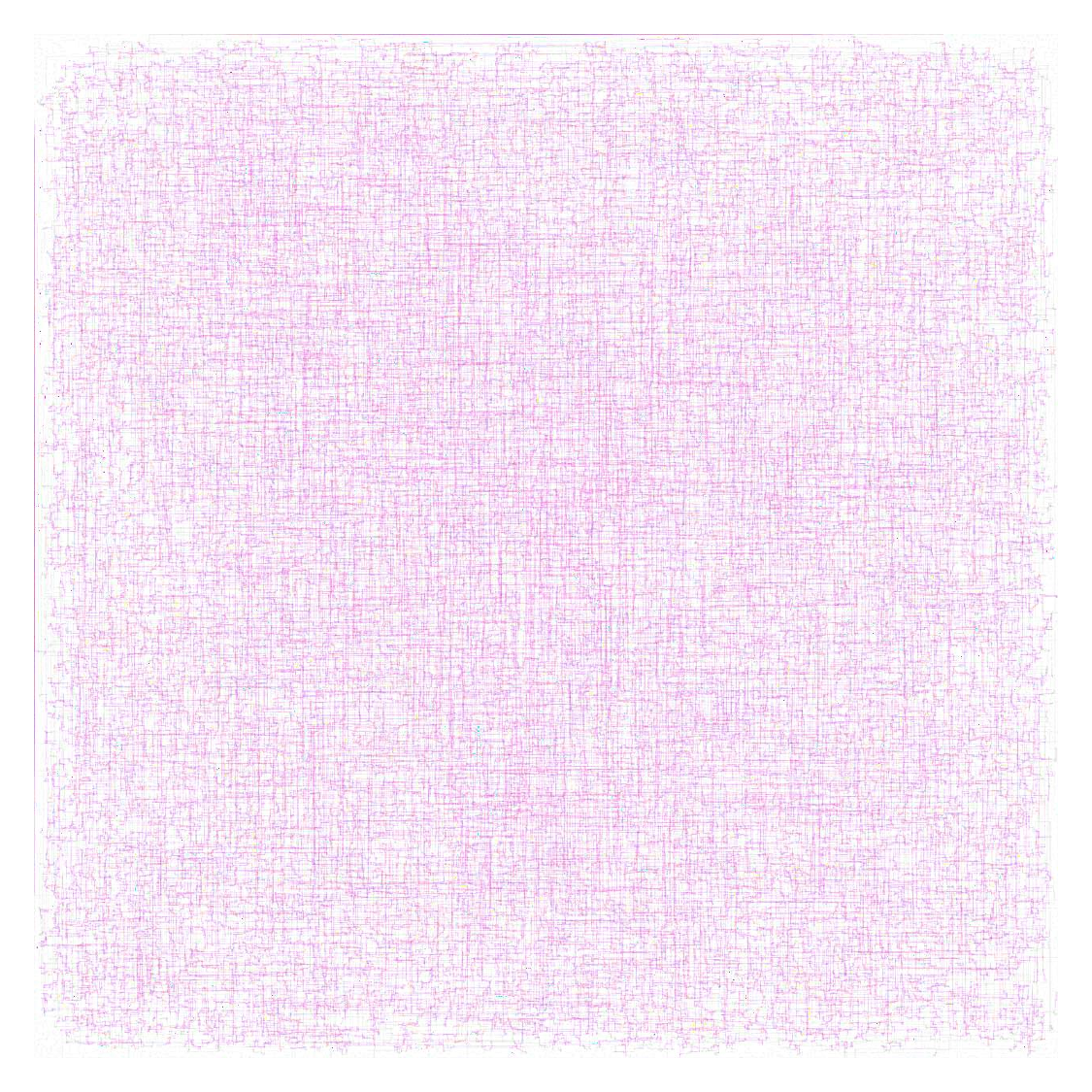}}
    \subfloat[Cir.4 with evenly sized partitions.] {\label{fig:visualization/riscv_parallel}\includegraphics[width=0.33\textwidth]{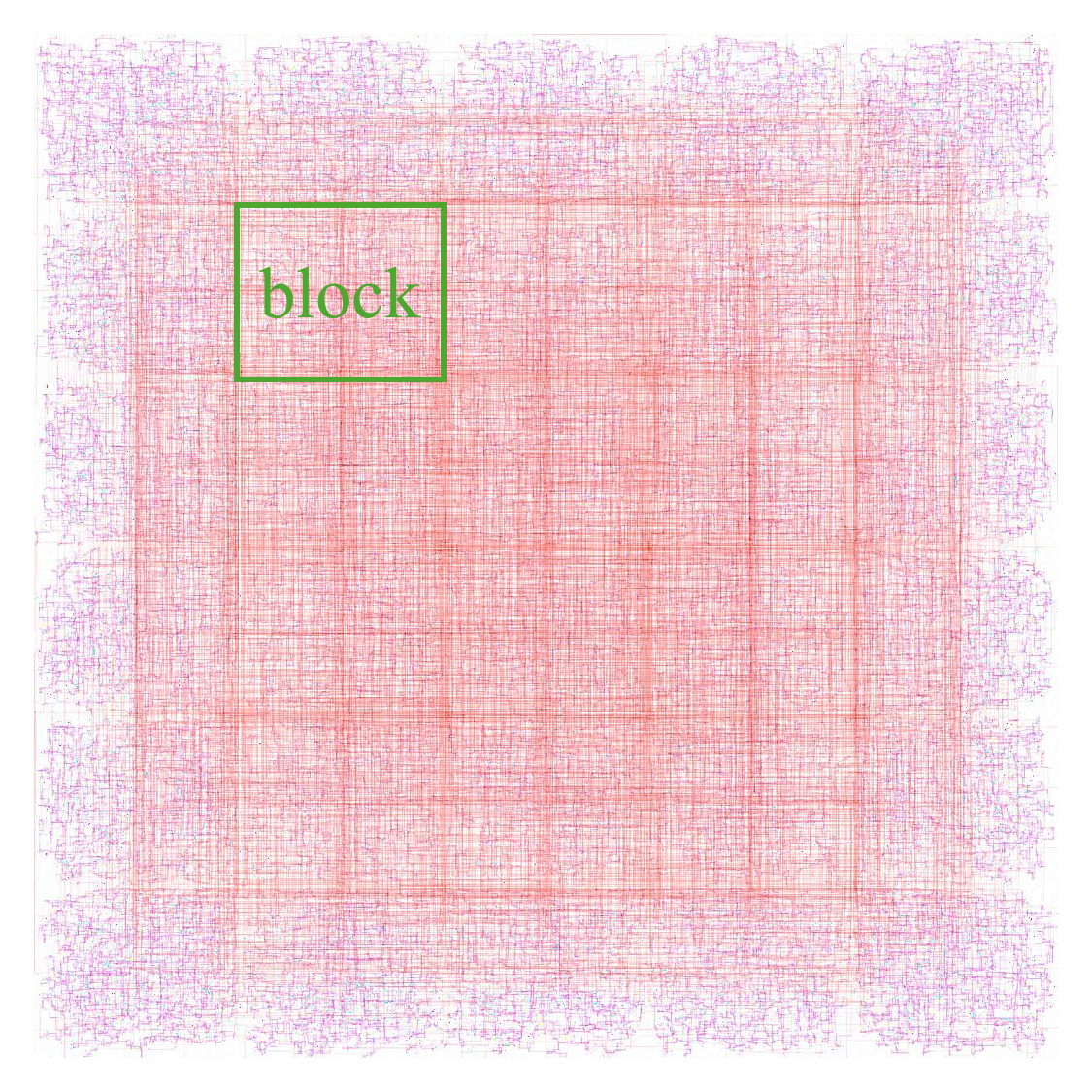}}
    \subfloat[Cir.4 pre-partitioned by semantics.] {\label{fig:visualization/riscv_manual}\includegraphics[width=0.33\textwidth]{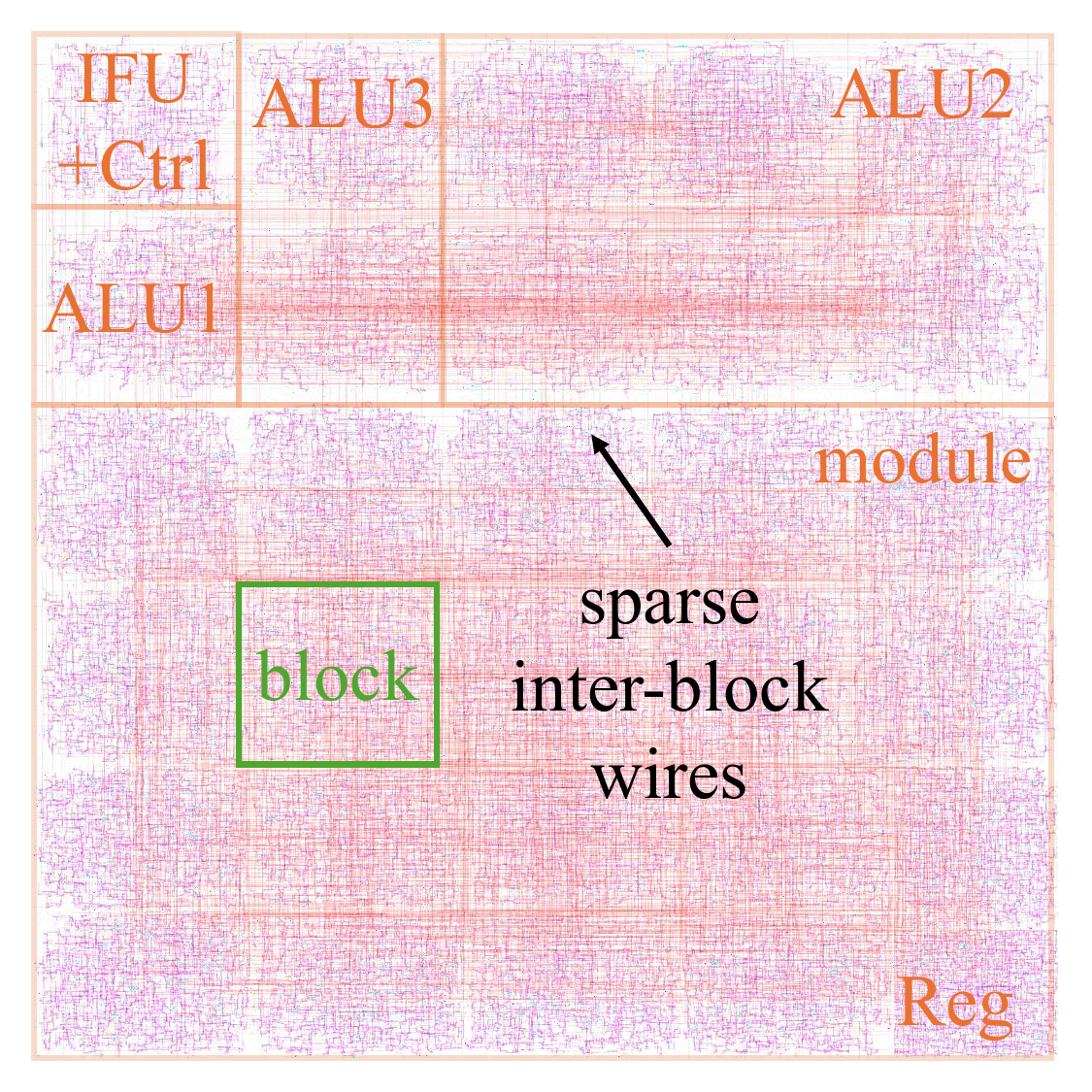}}
    \caption{Visualization of different approaches. Purple/Red lines: intra-block/inter-block wires.}
    \label{fig:visualization}
\end{figure*}

\subsection{Sensitivity Studies}
\label{sec:eval_breakdowns}

We further evaluate important individual parameters and optimizations in our approaches by conducting sensitivity studies on iteration number, partition number, virtual pin, and semantics.

\textbf{1) Iteration number:}
\cref{fig:evaluations_perceptron} shows the effect of iteration number on \CodeIn{Small} and \CodeIn{Large\&A*}, using Cir.2.
For \CodeIn{Large\&A*}, we evenly distribute the iterations to the four partitioned blocks. The iteration number per block is $T_N$.
In addition, we report $\Psi_r$ on \CodeIn{SA in Euclidean} and \CodeIn{SA in Manhattan} as a lower bound.
The runtime increases linearly, but \CodeIn{Large\&A*} is $4\times$ faster than \CodeIn{Small} because of processing four blocks in parallel (\cref{fig:evaluations/perceptron_runtime}).
Every additional iteration improves $\Psi_r$ (\cref{fig:evaluations/perceptron_twl}).
We observe that at about $T_N=N^2$, where $N$ is the component number for every block, 
$\Psi_r$ improvement hits a plateau of diminishing returns, thus indicating a sweet spot to stop iterating.
Reflected in \cref{fig:evaluations/perceptron_twl}, \CodeIn{Large\&A*} converges to a high-quality solution at the left vertical line.
The other three approaches converge at the right vertical line.
$\Omega$ is relative to $\Psi_r$, so the same observations apply (\cref{fig:evaluations/perceptron_insu}).

\smallskip

Next, we evaluate the effect of partition number, virtual pin, and semantics on approaches \CodeIn{Large\&A*} and \CodeIn{Large\&A*+Semantics}, using Cir.4.
For partition by semantics, we first partition Cir.4 into five modules, as shown in \cref{fig:visualization/riscv_manual}, and then partition it into a total of 30 blocks.
In each case, we evaluate the approaches with and without virtual pin.
We break down $\Psi_r$ into intra-block and inter-block, and $Rt$ into partition (circuit partitioning and floorplanning), intra-block (component placement and intra-block routing), and inter-block (inter-block routing).
\cref{fig:evaluations_riscv} shows our results.

\textbf{2) Partition number:}
Focusing on evenly sized partitions, we find the $Rt$ first decreases and then increases (\cref{fig:evaluations/riscv_runtime}), because, with few partitions, intra-block dominates the $Rt$, while with many partitions the partitioning logic itself dominates the $Rt$.
We identify the sweet spot of the number of partitions as 30--100 for Cir.4.
While the total $Rt$ in that range is almost constant, its breakdown shifts, with partitioning accounting for $12.8\%$ for 30 blocks increasing to $64.8\%$ for 100 blocks.
Consequently, the more instances of the same circuit one wants to manufacture, the larger the number of partitions to apply, if speed is important, as the overhead of partitioning must only be incurred once.
For example, using 64 partitions to manufacture one instance of Cir.4 is $1.35\times$ faster than using 100 partitions. In contrast, to manufacture 30 instances, using 100 partitions is $1.08\times$ faster.
Runtime aside, $\Psi_r$ keeps increasing with the partition number because of a growing length contribution from inter-block wires.
$\Omega$ is commensurate to $\Psi_r$.

\textbf{3) Virtual pin:}
By contrasting solid (off) and shaded (on) bars, we find that virtual pin reduces $\Psi_r$ (and consequently $\Omega$) in all cases, at the cost of a $6\%$ slowdown on average.
On average, $\Omega$ is $23.6\%$ lower and $\Psi_r$ is $15.3\%$ shorter, as a net result of a $14.6\%$ intra-block wire length increase and a $27.1\%$ inter-block wire length decrease.

\textbf{4) Semantics:}
By comparing shaded bars (i.e., virtual pin on) of 30 evenly sized blocks and partitioning using semantics, we find the intra-block wire length remaining almost the same.
However, leveraging semantics reduces inter-block wire length by $47\%$, resulting in $33\%$ lower $\Psi_r$ and $43\%$ less $\Omega$.

\subsection{Visualization}
\label{sec:eval_visual}

\cref{fig:visualization} depicts the effect of our approaches on circuit layout. 
\cref{fig:visualization/perceptron_single} shows an overview of Cir.2 without partitioning, with zoomed-in details in
\cref{fig:visualization/perceptron_detail}. %
Red circles denote insulators.
Green rectangles denote component bounding boxes.
The component at the bottom-right remains unused.
\cref{fig:visualization/perceptron_parallel} shows Cir.2 with four evenly sized partitions, where purple/red lines denote intra-/inter-block wires, respectively.
\cref{fig:visualization/riscv_single,fig:visualization/riscv_parallel,fig:visualization/riscv_manual} show Cir.4 without partitioning, with evenly sized partitions, and with partitioning by semantics, respectively.
The wires in \cref{fig:visualization/riscv_single} are evenly distributed across the whole space.
In \cref{fig:visualization/riscv_parallel}, intra-block wires are evenly distributed, but inter-block wires mainly travel on the boundaries and centers of blocks---a side-effect of virtual pin.
Components with pins connecting out of their block are navigated toward the boundaries of the block, rather than spread within the whole block's area.
This increases the density of wires at the edge and middle areas of each block, but is not a problem at the low circuit densities \pne can currently achieve.
Orange boxes in \cref{fig:visualization/riscv_manual} indicate the partitions created by leveraging semantic information directly extracted from Verilog modules.
The RISC-V core circuit is partitioned into IFU+Control, ALU1 (Add \& Sub \& Less Than), ALU2 (Mult), ALU3 (Others), and Register File.
A stripe with sparse inter-block wires is clearly discernible because, by the circuit's design, there are few connections between the modules.

\section{Conclusion and Future Work}
This paper explores the new tradeoff space introduced by nanomodular (and micromodular) electronics (\pne). 
To that end, we formulate and solve the placement and routing problems under the novel challenges of \pne, including imprecise component deposition.
Our workflow adopts standard physical design automation algorithms and adapts them to accommodate the technology's unique characteristics and balance all requirements.
Our results highlight the opportunity to accelerate component placement and wire routing by $952\times$ at a modest increase in wire length of $21\%$, together achieving a $108.3\times$ end-to-end manufacturing time improvement.
Our proposed workflow can be integrated into existing \pne manufacturing pipelines (e.g., \cite{10684980}), potentially accelerating the adoption of a technology that facilitates the democratization and rapid prototyping of custom electronics.

While our adapted approaches provide us with confidence in scaling \pne design automation to the target scale, there are still factors that are not considered in our work.
For example, we have not considered timing or power consumption requirements at this stage.
The balance between time-to-solution and the produced circuit quality can also be further fine-tuned as indicated in our evaluations.
We hope that addressing these limitations will encourage future works in \pne design automation.

\section*{Acknowledgments}

This work was supported by Speculative Technologies.


\bibliographystyle{abbrv}
\balance
\bibliography{references}

\end{document}